\documentclass[12pt]{elsarticle}
\usepackage{refmerge}
\usepackage{epsfig}
\graphicspath{{figs/}}
\begin{document}
\date{\today}

\title{Measurement of the $e^+e^- \to K^+K^-\pi^+\pi^-$ cross section with the CMD-3 detector at the VEPP-2000 collider}

\author[adr1,adr2]{D.N.Shemyakin\fnref{tnot}}
\author[adr1,adr2]{G.V.Fedotovich}
\author[adr1,adr2]{R.R.Akhmetshin}
\author[adr1,adr2]{A.N.Amirkhanov}
\author[adr1,adr2]{A.V.Anisenkov}
\author[adr1,adr2]{V.M.Aulchenko}
\author[adr1]{V.Sh.Banzarov}
\author[adr1]{N.S.Bashtovoy}
\author[adr1,adr2]{A.E.Bondar}
\author[adr1]{A.V.Bragin}
\author[adr1,adr2]{S.I.Eidelman}
\author[adr1,adr6]{D.A.Epifanov}
\author[adr1,adr2,adr3]{L.B.Epshteyn}
\author[adr1,adr2]{A.L.Erofeev}
\author[adr1,adr2]{S.E.Gayazov}
\author[adr1,adr2]{A.A.Grebenuk}
\author[adr1,adr2]{S.S.Gribanov}
\author[adr1,adr2,adr3]{D.N.Grigoriev}
\author[adr1,adr2]{F.V.Ignatov}
\author[adr1,adr2]{V.L.Ivanov}
\author[adr1]{S.V.Karpov}
\author[adr1,adr2]{V.F.Kazanin}
\author[adr1,adr2]{I.A.Koop}
\author[adr1,adr2]{A.A.Korobov}
\author[adr1,adr2]{O.A.Kovalenko}
\author[adr1,adr2]{A.N.Kozyrev}
\author[adr1,adr2]{E.A.Kozyrev}
\author[adr1,adr2]{P.P.Krokovny}
\author[adr1,adr2]{A.E.Kuzmenko}
\author[adr1,adr2]{A.S.Kuzmin}
\author[adr1,adr2]{I.B.Logashenko}
\author[adr1]{A.P.Lysenko}
\author[adr1,adr2]{P.A.Lukin}
\author[adr1,adr2]{K.Yu.Mikhailov}
\author[adr1]{V.S.Okhapkin}
\author[adr1]{Yu.N.Pestov}
\author[adr1,adr2]{E.A.Perevedentsev}
\author[adr1,adr2]{A.S.Popov}
\author[adr1,adr2]{G.P.Razuvaev}
\author[adr1,adr2]{Yu.A.Rogovsky}
\author[adr1,adr2]{A.A.Ruban}
\author[adr1]{N.M.Ryskulov}
\author[adr1,adr2]{A.E.Ryzhenenkov}
\author[adr1,adr2]{V.E.Shebalin}
\author[adr1,adr2]{B.A.Shwartz}
\author[adr1,adr2]{D.B.Shwartz}
\author[adr1,adr4]{A.L.Sibidanov}
\author[adr1]{Yu.M.Shatunov}
\author[adr1,adr2]{E.P.Solodov}
\author[adr1]{V.M.Titov}
\author[adr1,adr2]{A.A.Talyshev}
\author[adr1]{A.I.Vorobiov}
\author[adr1,adr2]{Yu.V.Yudin}
\author[adr1]{I.M. Zemlyansky}

\address[adr1]{Budker Institute of Nuclear Physics, SB RAS, 
Novosibirsk, 630090, Russia}
\address[adr2]{Novosibirsk State University, Novosibirsk, 630090, Russia}
\address[adr3]{Novosibirsk State Technical University, 
Novosibirsk, 630092, Russia}
\address[adr4]{University of Sydney, School of Physics, 
Falkiner High Energy Physics, NSW 2006, Sydney, Australia}
\address[adr6]{University of Tokyo, Department of Physics, 
7-3-1 Hongo Bunkyo-ku Tokyo, 113-0033, Japan}

\fntext[tnot]{Corresponding author:dimnsh@yandex.ru}

\begin{abstract}
\hspace*{\parindent}
The process $e^+e^- \to K^+K^-\pi^+\pi^-$ has been
studied in the center-of-mass energy range from 1500 to 2000\,MeV 
using a data sample of 23 pb$^{-1}$
collected with the CMD-3 detector 
at the VEPP-2000 $e^+e^-$ collider.
Using about 24000 selected events, the  $e^+e^- \to K^+K^-\pi^+\pi^-$ 
cross section has been measured with a systematic uncertainty
decreasing from 11.7\% at 1500-1600\,MeV to 6.1\% above 1800\,MeV.
A preliminary study of $K^+K^-\pi^+\pi^-$ production dynamics has 
been performed.

\end{abstract}

\maketitle

\section{ Introduction}
A high-precision measurement of the total cross section of
$e^{+}e^{-}\rightarrow~hadrons$ is important for various applications
including a calculation of the hadronic contribution to the muon anomalous 
magnetic moment $(g-2)_\mu$  in the frame of the Standard Model. To confirm
the existing difference between the calculated $(g-2)_\mu$ 
value~\cite{hagiwara} and the measured one~\cite{bnl}, new measurements of
the exclusive channels of $e^{+}e^{-}\rightarrow~hadrons$ with better 
accuracy are required. 
A contribution to the muon anomalous magnetic moment from  
$e^+ e^-\to K\bar K\pi\pi$ at the center-of-mass (c.m.) energies below 2\,GeV
is $(3.31 \pm 0.58) \times 10^{-10}$, estimated using isospin 
relations~\cite{hagiwara}. However, this result strongly depends on 
assumptions made about the presence of intermediate resonances, 
necessitating therefore a thorough study of the process dynamics in
various $K\bar K\pi\pi$ final states ($K^+K^-\pi^+\pi^-$, $K^+K^-\pi^0\pi^0$,  
$K^\pm K^0_{S(L)} \pi^\mp \pi^0$, $\ldots$).

The  process $e^+e^- \to K^+K^-\pi^+\pi^-$ has been earlier studied 
with the DM1~\cite{dm1} and DM2~\cite{dm2} detectors and more recently
with much larger effective integrated luminosity at the 
BaBar~\cite{babar1,babar2} and Belle~\cite{belle} detectors using 
an ISR approach. 
The first study of the production dynamics with the BaBar
detector exhibited a plethora of the resonant substructures 
($K^+K^-\rho$, $K^*K\pi$, $\phi\pi^+\pi^-$, $K_{1}K$ etc.) and some 
of them have been studied in more detail~\cite{babar2}.

In this paper we report a measurement of the $e^+e^- \to K^+K^-\pi^+\pi^-$ 
cross section and a preliminary study of production dynamics
based on 23 pb$^{-1}$ of an integrated luminosity collected 
by scanning of the c.m. energy ($E_{\rm c.m.}$) range from 1500 to 2000\,MeV. 

\section{Detector and data set}

The VEPP-2000 electron-positron collider~\cite{vepp2000} at Budker Institute
of Nuclear Physics (Novosibirsk, Russia) covers a c.m. energy range 
from 320 to 2000\,MeV and 
employs a novel technique of round beams
to reach luminosity up to 10$^{32}$\,cm$^{-2}$s$^{-1}$ at 
$E_{\rm c.m.}$=2000\,MeV.
The Cryogenic Magnetic Detector (CMD-3) described in~\cite{cmd3}
is installed in one of the two beam interaction regions.
The detector tracking system consists of the cylindrical drift chamber (DC) 
and double-layer cylindrical multiwire proportional Z-chamber, both also 
are used for a trigger and installed inside a thin (0.085 $X_{0}$)
superconducting solenoid with 1.3 T magnetic field.
DC contains 1218 hexagonal cells 
in 18 layers  and allows one to measure charged particle momentum 
with 1.5-4.5$\%$ accuracy in the (100-1000)\,MeV/$c$ range,
and the polar ($\theta$) and azimuthal ($\phi$) angles with 20 mrad and 
3.5-8.0 mrad accuracy, respectively.
Amplitude information from the DC wires is used to measure the 
ionization losses ($dE/dx_{\rm DC}$) of charged particles with $ {\sigma}_{dE/dx_{\rm DC}}=$11-14\% 
accuracy. The barrel electromagnetic calorimeters based on
liquid xenon (LXe) (5.4 $X_0$) and CsI crystals
(8.1 $X_{0}$) are placed outside the solenoid~\cite{CsI}.
The total amount of material in front of the calorimeter is 0.13 $X_{0}$ that includes the solenoid
as well as the radiation shield and vacuum vessel walls.
BGO crystals (13.4 $X_{0}$) are used as the endcap calorimeter. 
The flux return yoke is surrounded by scintillation counters
which are used to tag cosmic events.

To study a detector response and determine a detection efficiency,
we have developed a code for Monte Carlo (MC) 
simulation of our detector based on the GEANT4~\cite{GEANT4} package so that 
all simulated events are subjected to the same reconstruction and selection 
procedures as applied to the data. MC simulation of the signal process 
described further in Section~\ref{sec:mc} includes photon radiation by 
an initial electron or positron calculated according to~\cite{kur_fad}.

For the present analysis we use data of 2011 (1.0 T field) and 2012 
(1.3 T field) runs, collected at 66 beam energy points with 35 pb$^{-1}$ 
of an integrated luminosity.
In the 2011 run the energy range ($E_{\rm c.m.}$ = 1000--2000\,MeV) was scanned
up and down with  a 25\,MeV step collecting 
about 500\,nb$^{-1}$ per point.
In the 2012 run this range was scanned up with 20--40\,MeV steps 
collecting about 1\,pb$^{-1}$ per point. The integrated luminosity 
was determined using events of the processes 
$e^+e^- \to e^+e^-$ and $e^+e^- \to \gamma\gamma$
with about 1\% accuracy~\cite{lum}.

The beam energy was monitored by measuring the current 
in the dipole magnets of the main ring, and, in addition, at a few energy 
points by using the Back-Scattering-Laser-Light system~\cite{laser},~\cite{laser2}. 
Using the measured average momentum value of Bhabha events
and the average momentum of proton-antiproton pairs from 
the process $e^+ e^-\to p\bar p$~\cite{pp}, we determined $E_{\rm c.m.}$ for 
each energy point with about 6\,MeV and 2\,MeV accuracy for 2011 and 2012 
runs, respectively.  
\section{Events selection\label{sec:selections}}
Candidates for the events of the process under study are required to have 
three or four "good" tracks in the DC with the following "good" track 
definition:
\begin{itemize}
\item A track produces more than nine hits in the DC.
\item A track momentum is larger than 50\,MeV/$c$.
\item A minimum distance from the track to the beam axis in the 
transverse plane is less than 0.4\,cm.
\item A distance from the track to the center of the interaction 
region along the beam axis is less than 10\,cm.
\item A polar angle of the track is in the range from 0.85 to $\pi-$0.85 
radians. 
\item Ionization losses of the track are less than ionization losses 
of the proton.
\end{itemize}

\begin{figure}[hbtp]
\begin{minipage}[t]{0.46\textwidth}
	\centerline{\includegraphics[width=0.98\textwidth]{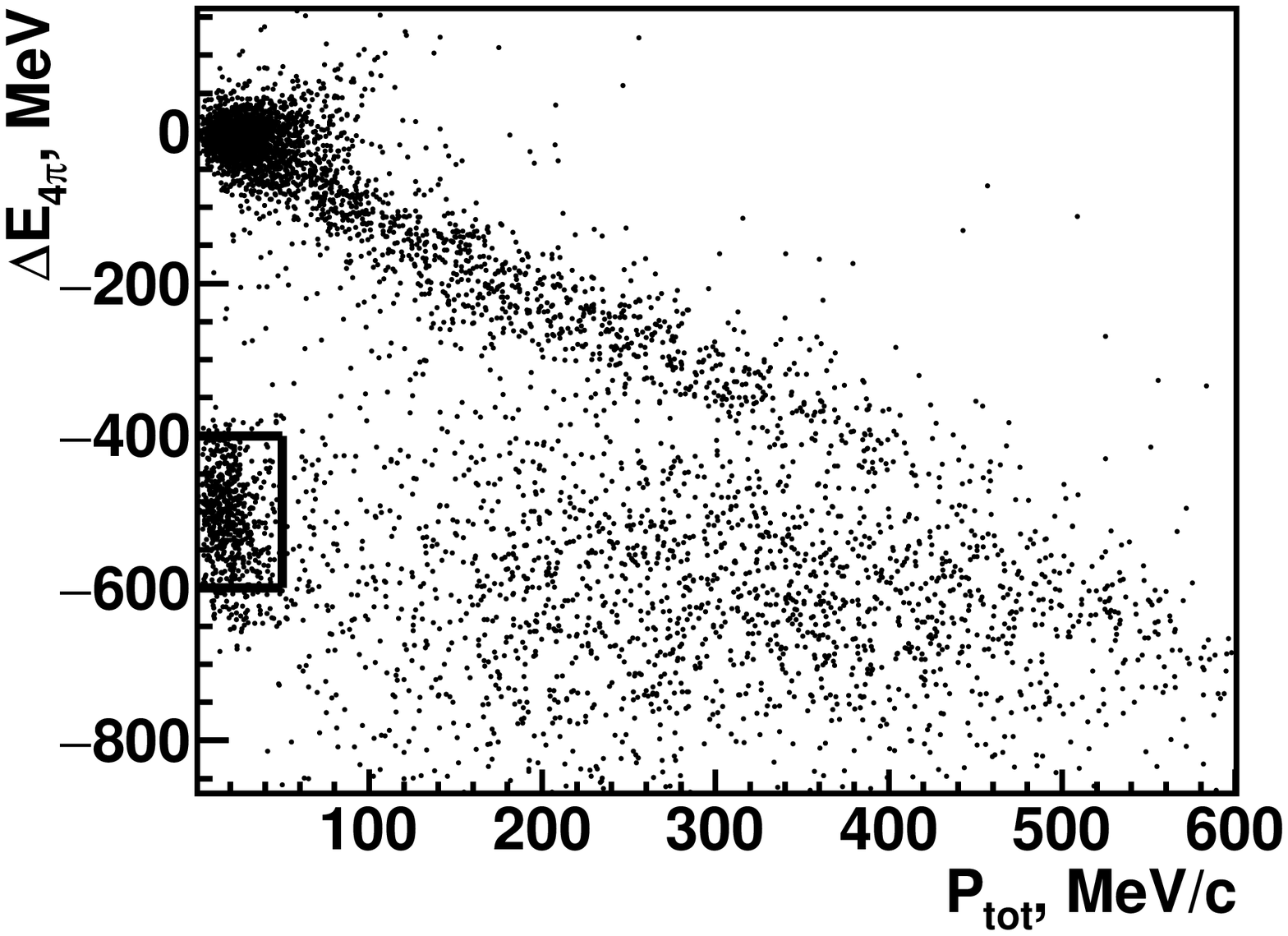}}
\caption
{Scatter plot of the $\Delta E_{4\pi}$ vs the total momentum  $P_{\rm tot}$ for 
four-track events at $E_{\rm c.m.}$=1980\,MeV.
\label{fig:e_p_4pi}}
\end{minipage}\hfill\hfill
\begin{minipage}[t]{0.46\textwidth}
	\centerline{\includegraphics[width=0.98\textwidth]{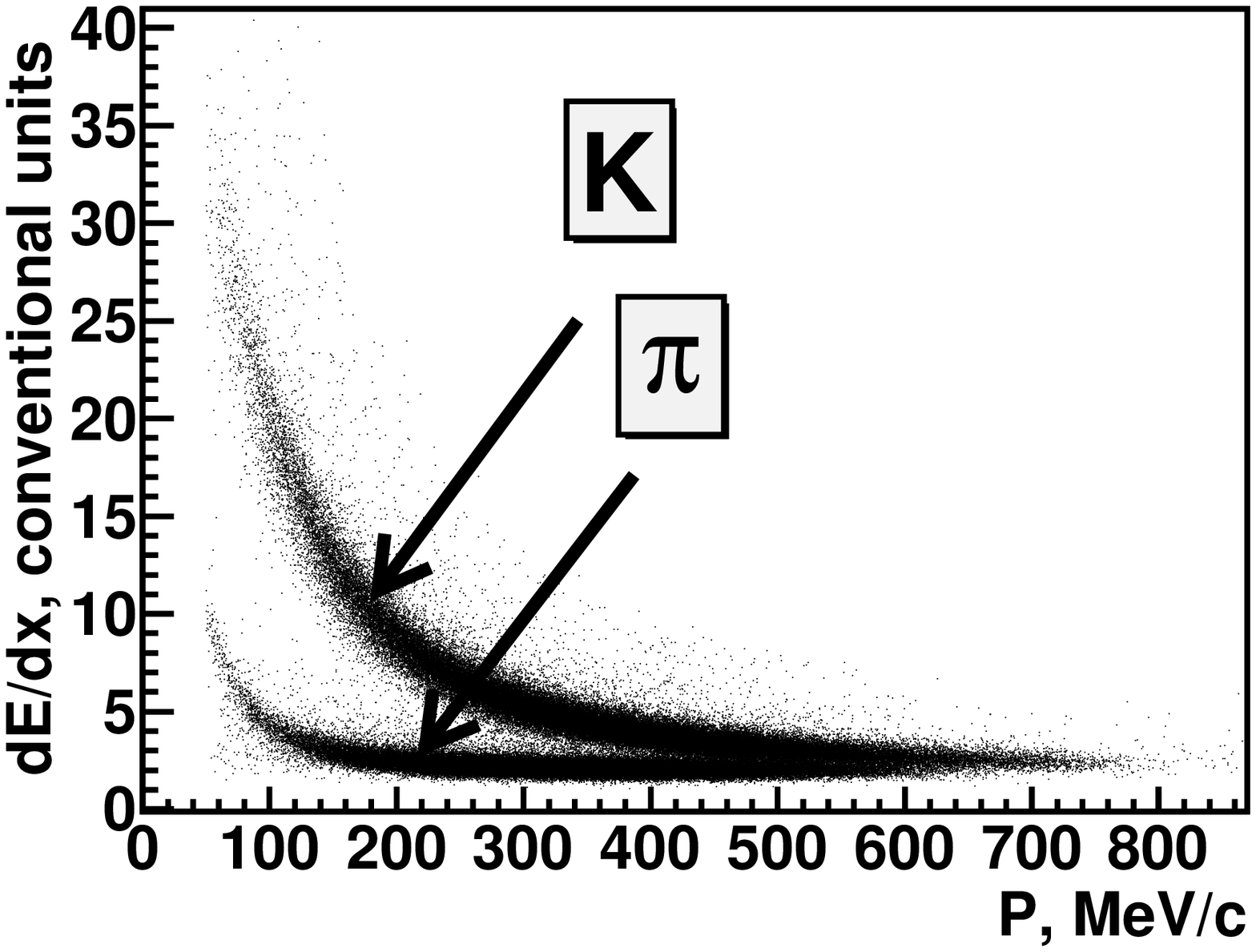}}
\caption
{The ionization losses in DC vs particle momentum for four-track events 
in simulation of signal process.
\label{fig:ion-los2d}}
\end{minipage}\hfill\hfill
\end{figure}

For selected events with four "good" tracks we calculate the total 
momentum $P_{\rm tot}$ and
the total energy $E_{4\pi}$, assuming all particles to be pions:
$$
 P_{\rm tot} =  | \sum_{i=1}^{4} \vec{p}_{i} |,~~~~~~~~~~E_{4\pi} = \sum_{i=1}^{4}\sqrt{p_{i}^2 + m_{\pi }^2}~.
$$
Figure~\ref{fig:e_p_4pi} shows a scatter plot of the difference 
$\Delta E_{4\pi} = E_{4\pi} - E_{\rm c.m.}$  vs the total momentum $P_{\rm tot}$.
The  $e^{+}e^{-} \to \pi^{+}\pi^{-}\pi^{+}\pi^{-}$ events locate near the 
origin of the coordinates.
Another cluster of events with a close to zero total momentum but shifted down along the 
vertical axis corresponds to $ K^{+}K^{-}\pi^{+}\pi^{-} $ events.
Events with high $P_{\rm tot}$ have missing particles and 
correspond to various background processes: 
$e^{+}e^{-} \to \pi^{+}\pi^{-}\pi^{+}\pi^{-}(\gamma)$,
$K_{S}^0 K^\pm\pi^\mp(\gamma)$, 
$\pi^{+}\pi^{-}\pi^{+}\pi^{-}\pi^{0}(\gamma)$,
$\pi^{+}\pi^{-}\pi^{+}\pi^{-}\pi^{0}\pi^{0}(\gamma)$, 
$\pi^{+}\pi^{-}\pi^{+}\pi^{-}\pi^{+}\pi^{-}(\gamma)$.

Using the selections $P_{\rm tot} < $ 50\,MeV$/c$ and 
$| \Delta E_{4\pi} + 500| < $ 100\,MeV 
we obtain a sample of $ K^{+}K^{-}\pi^{+}\pi^{-} $ events with a 
contribution from the background processes of  less than 5\% estimated 
according to simulation.
These events are used to develop a procedure of separation of pions and kaons.

\begin{figure}[tbh]
\begin{center}
	\includegraphics[width=0.9\textwidth]{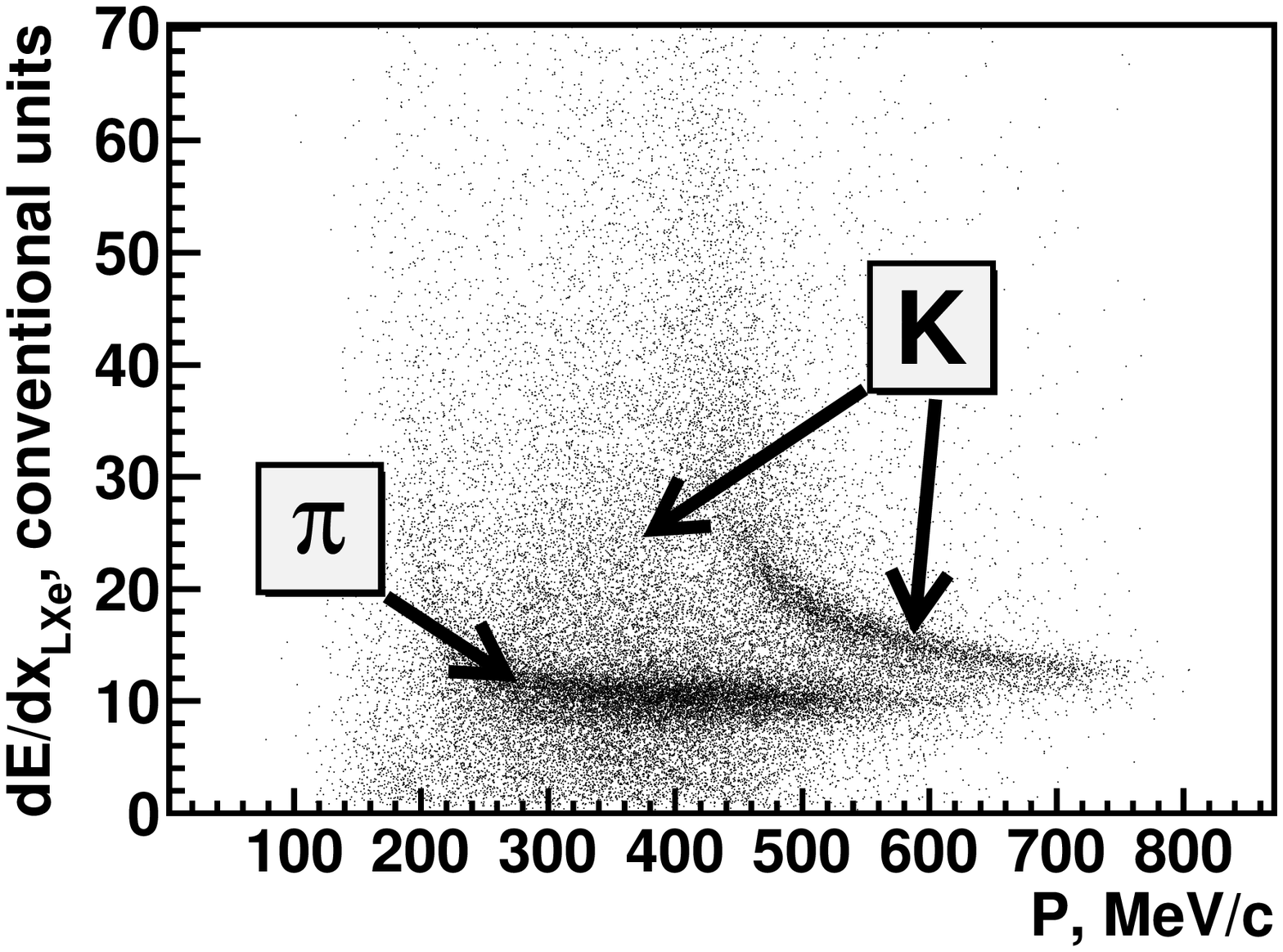}
\caption
{Ionization losses in the LXe calorimeter vs momentum for 
$ K^{+}K^{-}\pi^{+}\pi^{-} $ events in simulation.}
\label{fig:ion-los2d_lxe}
\end{center}
\end{figure}

$K/\pi$ separation in CMD-3 can be performed by analysing ionization losses in
the DC and the LXe calorimeter. The ionization losses 
$dE/dx_{\rm DC}$ in the DC
for kaons and pions from simulated $ K^{+}K^{-}\pi^{+}\pi^{-}$ events are shown 
in Fig.~\ref{fig:ion-los2d}.
A similar plot for the ionization losses $dE/dx_{\rm LXe}$ in 
the LXe calorimeter is shown in Fig.~\ref{fig:ion-los2d_lxe}.
It can be seen that $dE/dx_{\rm DC}$ differ significantly for kaons and pions 
for momenta less than 500\,MeV/$c$, while $dE/dx_{\rm LXe}$ differ even 
at higher momenta,
so one can perform $K/\pi$ separation in the whole particle momentum range. 
The distributions of the ionization losses have different shapes in
the DC and the LXe calorimeter.
It is Gaussian for $dE/dx_{\rm DC}$, while nuclear interactions are likely 
in the LXe calorimeter, resulting in the wide tails of the $dE/dx_{\rm LXe}$ 
distribution, in which about 20\% of the events are located.
Simulation of nuclear interactions of kaons and pions in LXe is not perfect,
especially at low momenta, so using simulated $dE/dx_{\rm LXe}$ results 
in uncontrollable systematic uncertainties.
Thus an experimental input is required for studying energy deposition of 
particles in the LXe calorimeter and ionization losses in the drift 
chamber are only used in this analysis. This results in a
less than 0.5\% uncertainty in $K/\pi$ separation estimated from simulation,
where the particle type and the  number of misidentified particles are known.
This uncertainty is negligible
compared to the total systematic error (discussed in section~\ref{sec:systematic}).

\begin{figure}[hbtp]
\begin{minipage}[t]{0.46\textwidth}
	\centerline{\includegraphics[width=0.98\textwidth]{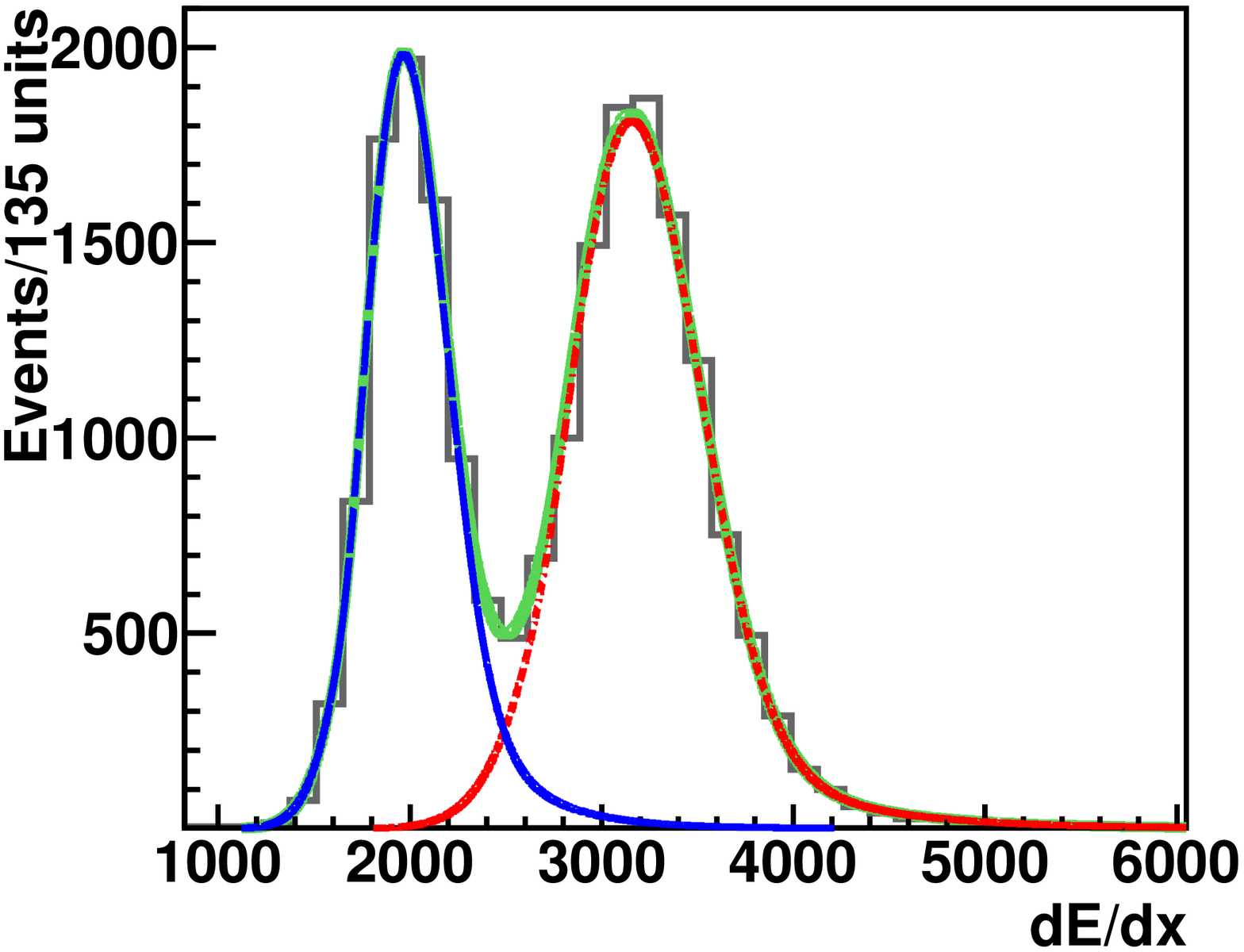}}
\caption
{Ionization losses for tracks with momentum in the 470-520\,MeV/$c$ range. 
The lines show PDFs for pions and kaons, described in the text.
\label{fig:pdf}}
\end{minipage}\hfill\hfill
\begin{minipage}[t]{0.46\textwidth}
	\centerline{\includegraphics[width=0.98\textwidth]{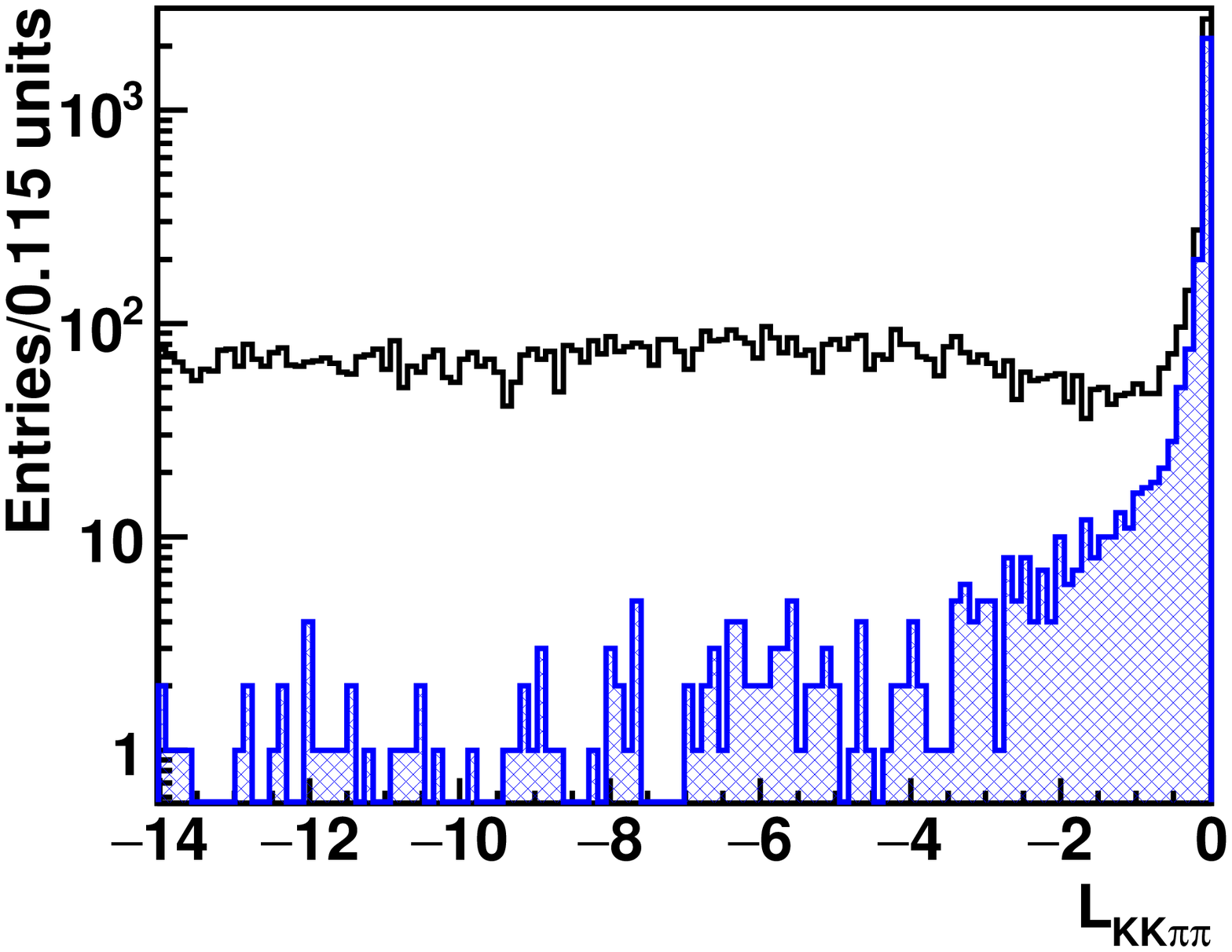}}
\caption
{Distribution of the likelihood function $L_{KK\pi\pi}$ for all three- 
and four-track events in data (open histogram), and
for $K^+K^-\pi^+\pi^- $ simulated events (hatched histogram).
\label{fig:LHF}}
\end{minipage}\hfill\hfill
\end{figure}

We use momentum and $dE/dx_{\rm DC}$ values for each track to construct  
probability density functions (PDF) for kaons $f_{K}(p, dE/dx_{\rm DC})$ and 
pions $f_{\pi}(p, dE/dx_{\rm DC})$, each of which is a sum of Gaussian and 
logarithmic Gaussian distributions.
The parameters of PDF are determined by 
approximating the $dE/dx_{\rm DC}$ histogram by PDF. 
The parameters of these functions are fitted by smooth lines
which depend on momentum.
First we use a sample of $e^+e^- \to \pi^+\pi^-\pi^+\pi^-$ events to 
determine $f_{\pi}(p, dE/dx_{\rm DC})$, then the function 
$f_{K}(p, dE/dx_{\rm DC})$ is 
determined using $e^+e^- \to K^+K^-\pi^+\pi^-$ events.
The control sample of $e^+e^- \to \pi^+\pi^-\pi^+\pi^-$ events is 
selected using strong cuts, that reject the background to the level of
less than 0.5\%.
The background in the sample of $e^+e^- \to K^+K^-\pi^+\pi^-$ events is 
about 1\% and will discussed below in this section.
This procedure is performed separately for simulation and experiment. 
The approximation of the simulated $dE/dx_{\rm DC}$ distribution with a sum of two functions 
$f_{K}(p, dE/dx_{\rm DC})$ and $f_{\pi}(p, dE/dx_{\rm DC})$ in the momentum range 
470--520\,MeV/$c$ is presented in Fig.~\ref{fig:pdf}.

Selection of events of the process $e^+e^- \to K^+K^-\pi^+\pi^-$ from 
the three- and four-track samples is performed using a likelihood function 
$L_{KK\pi\pi}$, which is constructed as:
$$
L_{KK\pi\pi} =    \ln \left( {\frac{{\prod f_{\alpha}^{i} (p, dE/dx_{\rm DC})}}{\prod [f_{\pi}^{i} (p, dE/dx_{\rm DC}) + f_{K}^{i}(p, dE/dx_{\rm DC})]}} \right),
$$
where $i$ --- track index, which varies from 1 to 3 or 4,
$\alpha_{i} = K,\pi$ --- assumed type of a particle corresponding to the 
$i$-th track. Under the assumption that each event is from  the process 
$e^+e^- \to K^+K^-\pi^+\pi^-$, two tracks with opposite charge should be 
identified as kaons while the other two as pions and we test all possible 
combinations of $\alpha_{i} $ to obtain the maximum value of $L_{KK\pi\pi}$.

Figure~\ref{fig:LHF} shows the $L_{KK\pi\pi}$ value for  all three- and 
four-track events in data by the open histogram, while the hatched histogram
corresponds to the $K^+K^-\pi^+\pi^- $ events from simulation.
The likelihood function value is a good parameter for separating 
$K^+K^-\pi^+\pi^-$ events from the background. We applied 
a requirement  $L_{KK\pi\pi} > -3$ which retains more than 95\% of signal events.

\begin{figure}[hbtp]
\begin{minipage}[t]{0.46\textwidth}
	\centerline{\includegraphics[width=0.98\textwidth]{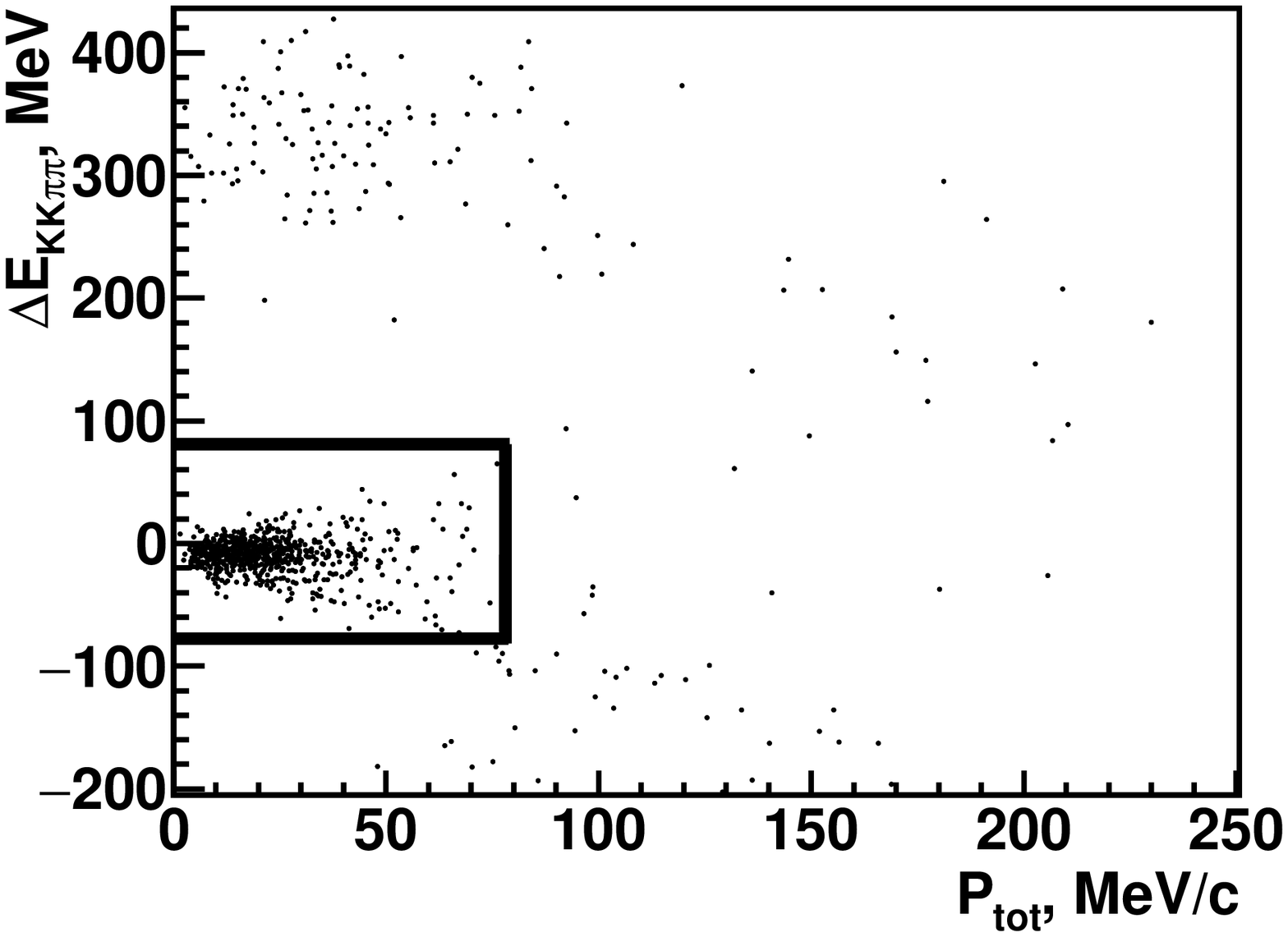}}
\caption
{The scatter plot of the difference $\Delta E_{KK\pi\pi}$  
between the total energy and c.m. energy vs the total momentum for 
the four-track events at $E_{\rm c.m.}$=1980\,MeV.  
The rectangle shows the selected area.
\label{fig:4e_p}}
\end{minipage}\hfill\hfill
\begin{minipage}[t]{0.46\textwidth}
	\centerline{\includegraphics[width=0.98\textwidth]{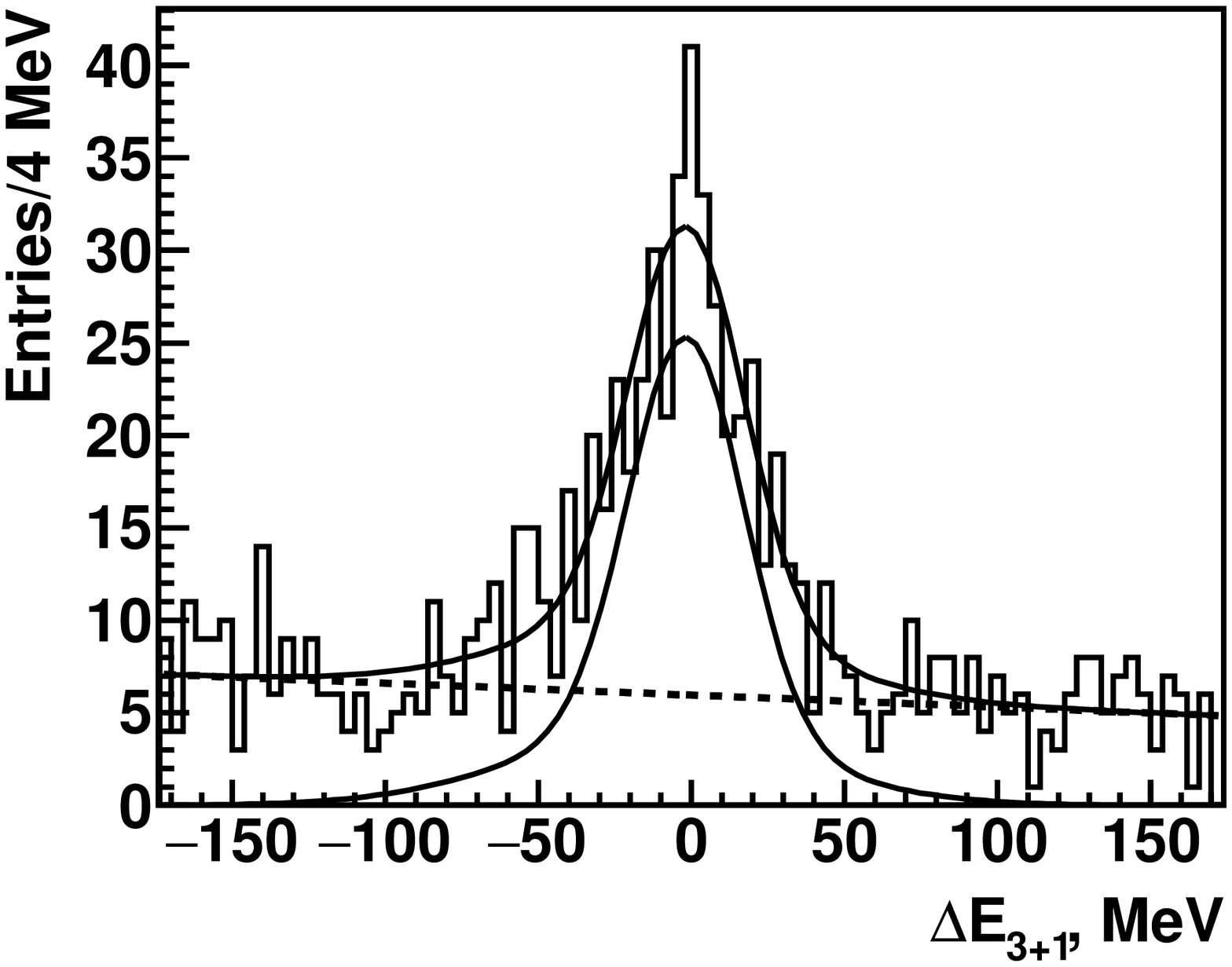}}
\caption
{Difference $\Delta E_{3+1}$ between the calculated 
	total energy for events with one missing particle and the c.m. energy at $E_{\rm c.m.}$=1980\,MeV.  
The lines show a fit explained in the text. 
\label{fig:3energy_fl_p}}
\end{minipage}\hfill\hfill
\end{figure}

We assign pion or kaon mass to each track and calculate total energy 
$E_{KK\pi\pi}$. Figure~\ref{fig:4e_p} shows a scatter plot of the difference 
$\Delta E_{KK\pi\pi} = E_{KK\pi\pi} - E_{\rm c.m.}$ vs
the total momentum for events with four tracks assuming all events to be 
$e^+e^- \to K^+K^-\pi^+\pi^-$, and the condition $L_{KK\pi\pi} > -3$ 
applied. Events of the process  $e^+e^- \to K^+K^-\pi^+\pi^-$ are located 
near the origin of the coordinates.

The width of the $E_{KK\pi\pi}$ distribution for  $K^{+}K^{-}\pi^{+}\pi^{-}$ events
is a few times smaller than that of the $E_{4\pi}$ distribution while the 
shapes of the background 
distribution are the same for $E_{4\pi}$ and $E_{KK\pi\pi}$.
For that reason, conditions on $\Delta E_{KK\pi\pi}$ and $P_{\rm tot}$
allow to suppress the $e^{+}e^{-} \to \pi^{+}\pi^{-}\pi^{+}\pi^{-}(\gamma)$ 
background. However, five- and six-body processes with a missing particle 
in the final state ($\pi^{+}\pi^{-}\pi^{+}\pi^{-}\pi^{0}(\gamma)$,
$\pi^{+}\pi^{-}\pi^{+}\pi^{-}\pi^{0}\pi^{0}(\gamma)$, 
$\pi^{+}\pi^{-}\pi^{+}\pi^{-}\pi^{+}\pi^{-}(\gamma)$), or four-body processes 
with charged kaons 
($K_{S}^0 K^\pm\pi^\mp(\gamma)$ at c.m. energy less than 2\,GeV)
have the same value of $E_{KK\pi\pi}$ and $ P_{\rm tot}$ as the signal process.
The condition on $L_{KK\pi\pi}$ allows one to significantly reduce the number 
of such background events.
Application of the requirements $|\Delta E_{KK\pi\pi}| < 80$\,MeV, 
$ P_{\rm tot} < 80$\,MeV/$c$ and $L_{KK\pi\pi} > -3$
decreased  the background to less than 1\% level as estimated using
the multihadron Monte Carlo generator~\cite{mhg}.
In this generator all experimentally measured processes of $e^+e^-$ 
annihilation into hadrons up to 2\,GeV are used to calculate a total cross 
section at a given c.m. energy and events of each final state are
sampled with a probability proportional to the  fraction of its measured 
cross section in the total one.

The  $L_{KK\pi\pi} > -3$ requirement for the three-track events selects 
candidates for the $K^{+}K^{-}\pi^{+}\pi^{-}$ final state for which  we know 
the charge and type of a missing particle.
Using the total momentum of three detected tracks and momentum 
conservation, we calculate the energy 
of the missing particle and add it to the energy of the detected particles 
to obtain the total energy $E_{3+1}$ for these events.
The histogram in Fig.~\ref{fig:3energy_fl_p} shows the difference between 
the calculated total energy for events with one missing particle  and the
c.m. energy, $\Delta E_{3+1} = E_{3+1} - E_{\rm c.m.}$. 
To extract the yield of signal events, we fit this distribution with a 
signal function and a linear background.
The signal function shape is obtained using four-track $K^+K^-\pi^+\pi^-$ 
events assuming that one track is not detected.
According to simulation~\cite{mhg}, the linear function describes 
the background well.
 
Events with a missing particle mainly appear due to the limited DC acceptance, 
in addition, some tracks are not reconstructed due to a DC inefficiency,
decays in flight or nuclear interaction.

Using distributions of $\Delta E_{3+1}$ parameter we determine separately 
the number of missing pions or kaons expected inside and outside the DC 
acceptance. Events with a missing track are used to study a reconstruction 
efficiency as discussed in Section~\ref{sec:efficiency}.

In total, 10545 four-track and 13349 three-track events of the  
process $e^+e^- \to K^+K^-\pi^+\pi^-$ have been selected. 
The numbers of signal events at each energy point are listed in Table 1.

\section{Simulation\label{sec:mc}}
A MC generator of the process  $e^+ e^-\to K^+K^-\pi^+\pi^-$ has been developed
to obtain a detector response to $K^+K^-\pi^+\pi^-$ events and to calculate 
the detection efficiency.
Since the DC acceptance is only 70\% of the total solid angle,
the correct determination of the total detection efficiency 
requires adequate simulation of the production dynamics of $e^+ e^-\to K^+K^-\pi^+\pi^-$ events. 
The BaBar Collaboration reported observation of the following intermediate 
states for this process~\cite{babar2}:
$\phi(1020)f_0(980)$, $\phi(1020)f_0(500)$, $K_1(1270,1400) K$, 
$K^*(892)^0 K\pi$, $K^*_2(1430)^0 K$ and $\rho(770)K^+K^-$.
Following this study, we developed a MC generator which includes various
processes resulting in the $K\bar{K}\pi\pi$ final state:
\begin{itemize}
\item $e^{+}e^{-} \to K^*(892)^0 K\pi + c.c.$
\item $e^{+}e^{-} \to K^*(892)^0 \bar{K}^*(892)^0$
\item $e^{+}e^{-} \to f_0(980)\phi$
\item $e^{+}e^{-} \to f_0(500) \phi$
\item $e^{+}e^{-} \to \rho (KK)_{\rm S-wave}$
\item $e^{+}e^{-} \to (K_1(1270) K)_{\rm S-wave} \to (K^* \pi)_{\rm S-wave} K$
\item $e^{+}e^{-} \to (K_1(1400) K)_{\rm S-wave} \to (K^* \pi)_{\rm S-wave} K$
\item $e^{+}e^{-} \to (K_1(1270) K)_{\rm S-wave} \to (\rho K)_{\rm S-wave}K$
\end{itemize}

The matrix element 
$M(\vec p_{K^+},\vec p_{K^-},\vec p_{\pi^-},\vec p_{\pi^+}, \vec \alpha)$ 
is written as a weighted sum of the amplitudes of all intermediate states 
mentioned above with the relative phases assumed to be $0^\circ$ or $180^\circ$. 
The propagators of the intermediate resonances include relevant energy 
dependence of the width.
The weights of the intermediate-state amplitudes are obtained from a 
minimization of the likelihood function
$$L(\vec \alpha)=\prod_{i} \frac{P^i_{\rm det}(\vec \alpha)}{Z(\vec \alpha)}, $$
where multiplication is performed for experimental events, 
$P_{\rm det}(\vec \alpha)$ --- the probability to detect an 
$e^+ e^-\to K^+K^-\pi^+\pi^-$ event,
$\vec \alpha$ is the vector of the weights -- free parameters of the fit,
$Z(\vec \alpha)$ --- normalization coefficient.
The probability to detect an $e^+ e^-\to K^+K^-\pi^+\pi^-$ event 
with particle momenta 
$(\vec p_{K^+},\vec p_{K^-},\vec p_{\pi^-},\vec p_{\pi^+})$ is defined as:
$$P^i_{\rm det} = |M(\vec p^i_{K^+},\vec p^i_{K^-},\vec p^i_{\pi^-},\vec p^i_{\pi^+}, \vec \alpha)|^2 \epsilon(\vec p^i_{K^+},\vec p^i_{K^-},\vec p^i_{\pi^-},\vec p^i_{\pi^+}) \Delta F^i, $$
where $\epsilon(\vec p^i_{K^+},\vec p^i_{K^-},\vec p^i_{\pi^-},\vec p^i_{\pi^+})$ --- detection efficiency, $\Delta F$ --- element of the event phase space.
The calculation of 
$$Z(\vec \alpha) = \frac{\int |M(\vec p_{K^+},\vec p_{K^-},\vec p_{\pi^-},\vec p_{\pi^+}, \vec \alpha)|^2 \epsilon(\vec p_{K^+},\vec p_{K^-},\vec p_{\pi^-},\vec p_{\pi^+}) dF}{N_{\rm event}},$$
where $N_{\rm event}$ -- the total number of events, is performed using 
MC simulation.
It can be seen that 
$$L \propto \prod_{i} \frac{|M(\vec p^i_{K^+},\vec p^i_{K^-},\vec p^i_{\pi^-},\vec p^i_{\pi^+}, \vec \alpha)|^2}{Z(\vec \alpha)}$$
for a fixed sample of events, because the factor $\prod_{i} \epsilon(\vec p^i_{K^+},\vec p^i_{K^-},\vec p^i_{\pi^-},\vec p^i_{\pi^+}) \Delta F^i$ is independent of $\vec \alpha$.
The last equation for $L$ was used in minimization.

Since the number of events at each c.m. energy point is not sufficient 
to determine the weights, the events are combined into nine groups with the
following average values of $E_{\rm c.m.}$: 1600, 1650, 1700, 1750, 1800, 1850, 
1900, 1950 and 2000\,MeV.

We found no significant contribution from the $K^*(892)^0 K\pi$  and 
$K^*(892)^0 \overline{K}^*(892)^{0}$ channels and excluded them from $L$.
We observe strong dependence of the weights of the intermediate-state 
amplitudes on the c.m. energy and high interference between various 
amplitudes. It was found that the main intermediate mechanisms are 
$e^{+}e^{-} \to (K_1(1270, 1400) K)_{\rm S-wave} \to (K^* \pi)_{\rm S-wave} K$,
their contribution is about 50-90\% and depends on c.m. energy.
The $K_1(1270)$ and $K_1(1400)$ are broad resonances and phase space of 
the $K_1K$ system is small at energies smaller than 2\,GeV.
So a much larger sample is needed for separation of the intermediate states 
with these particles.

The dependence of the weights of the intermediate-state amplitudes on the 
c.m. energy has been smoothed and then used to construct a MC generator
that was used for the final detection efficiency calculation.
The obtained contributions of various intermediate states
are consistent with the BaBar results.
A detailed analysis of the production dynamics will be performed 
after increasing statistics and a study of the other charge combinations
of the $K\bar{K}\pi\pi$ final state.
\begin{figure}[hbtp]
\begin{minipage}[t]{0.46\textwidth}
	\centerline{\includegraphics[width=0.98\textwidth]{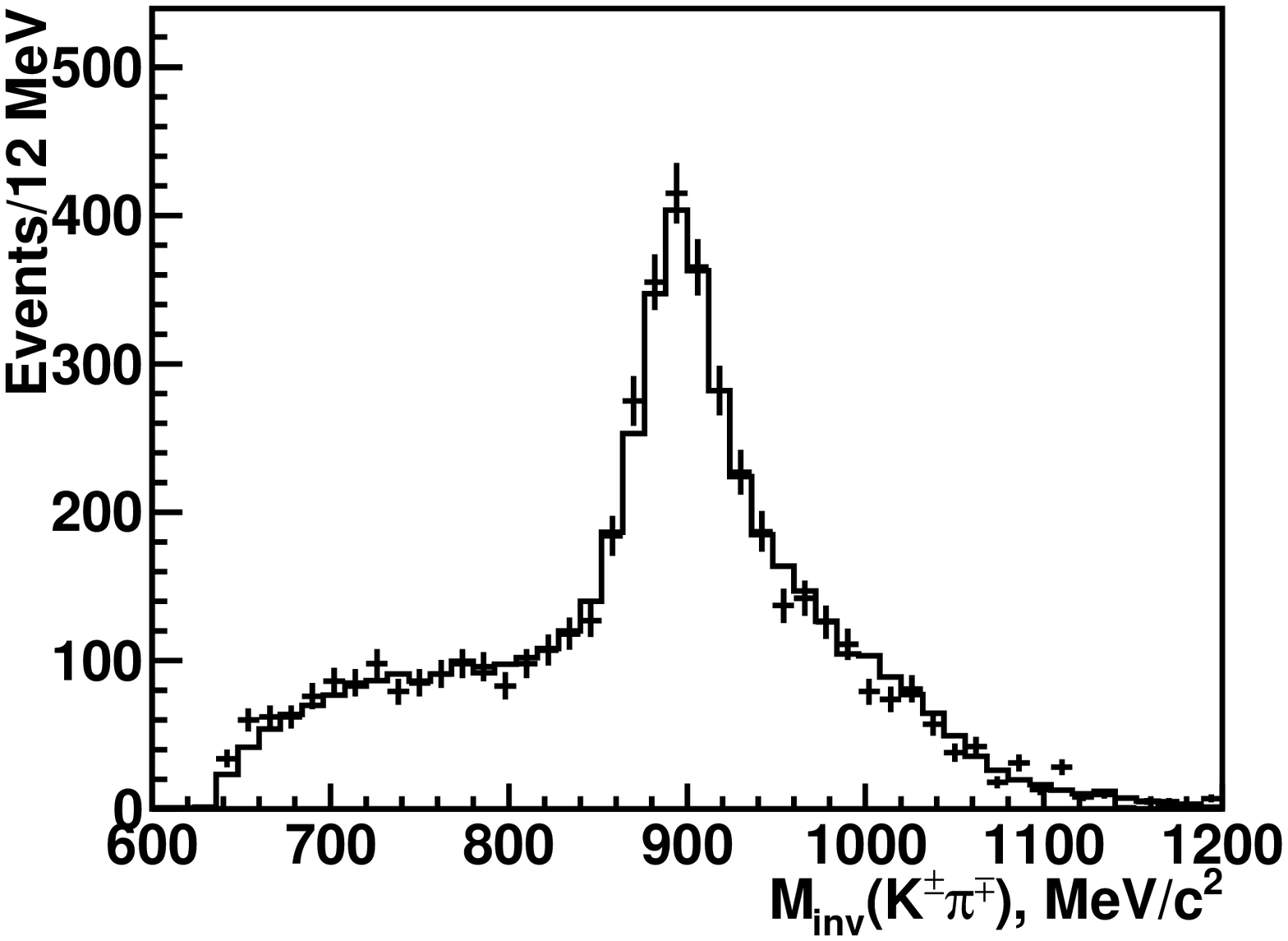}}
\caption
{$K^{\pm} \pi^{\mp}$ invariant mass for data (points) and MC simulation 
(histogram) at $E_{\rm c.m.}=1950$~MeV.
\label{fig:m_kpi_mc_exp}}
\end{minipage}\hfill\hfill
\begin{minipage}[t]{0.46\textwidth}
	\centerline{\includegraphics[width=0.98\textwidth]{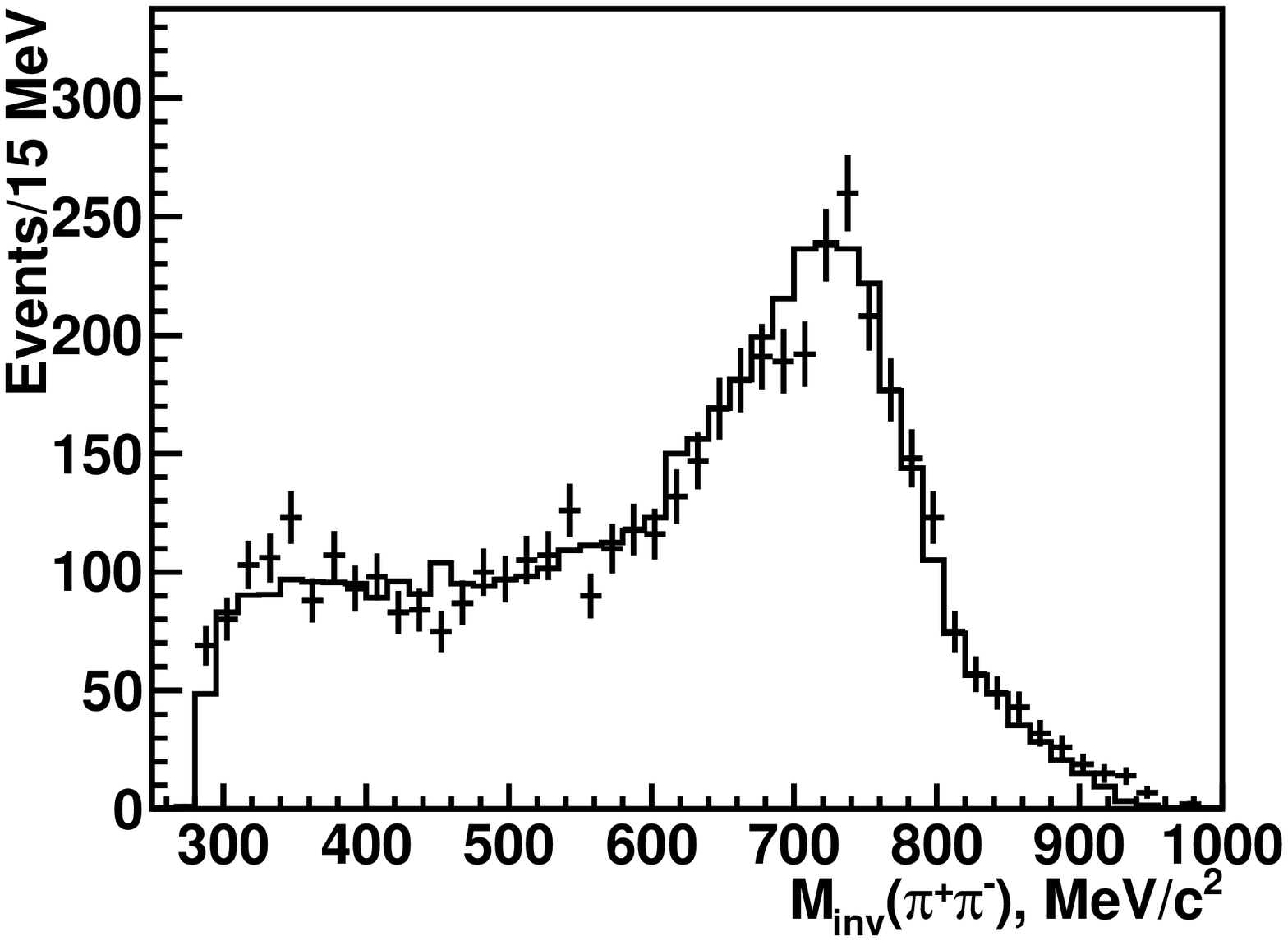}}
\caption
{$\pi^+\pi^-$ invariant mass for data (points) and MC simulation 
(histogram) at $E_{\rm c.m.}=1950$~MeV.
\label{fig:m_pipi_mc_exp}}
\end{minipage}\hfill\hfill
\end{figure}
\begin{figure}[hbtp]
\begin{minipage}[t]{0.46\textwidth}
	\centerline{\includegraphics[width=0.98\textwidth]{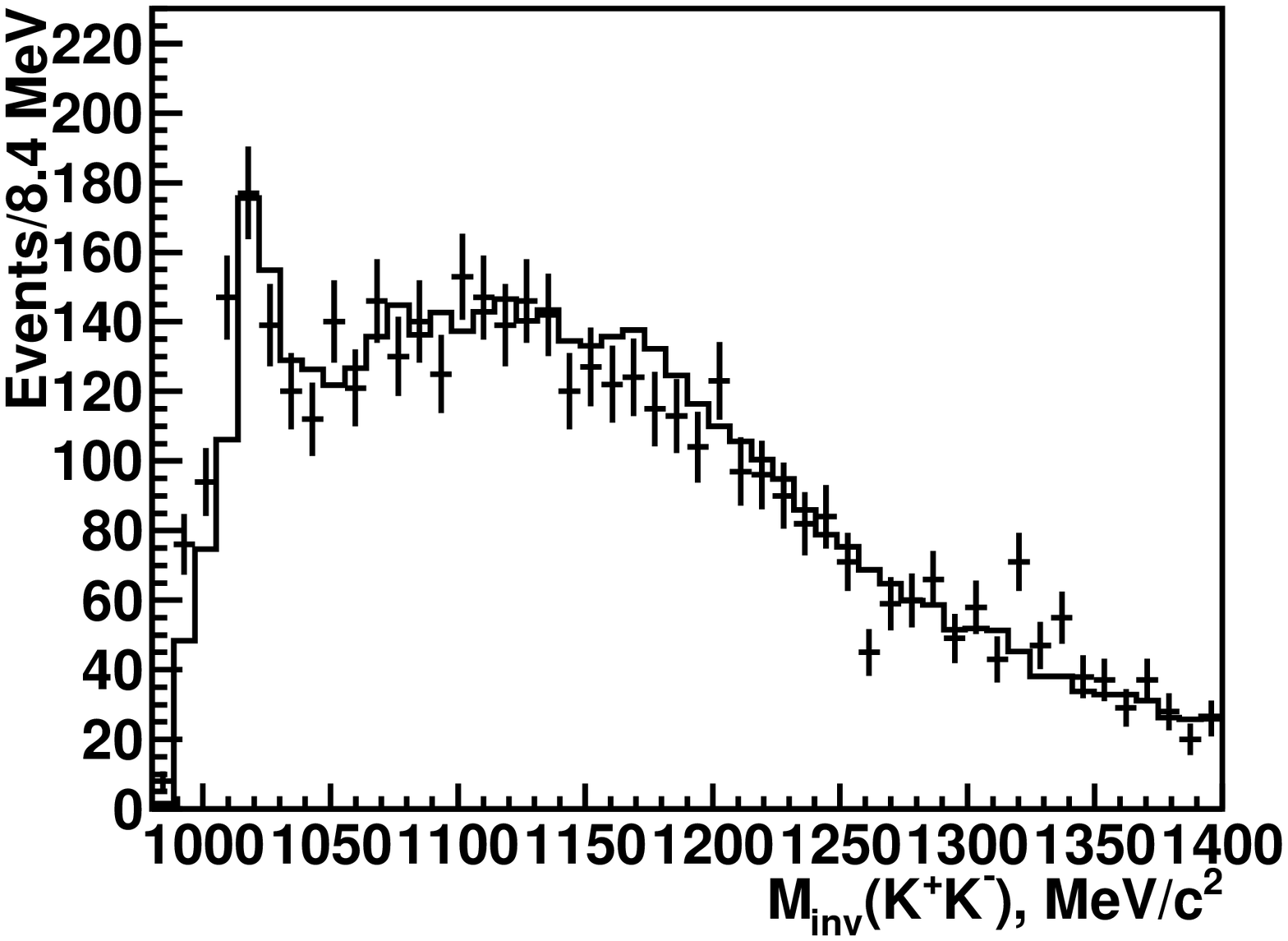}}
\caption
{$K^+ K^-$ invariant mass for data (points) and MC simulation 
(histogram) at $E_{\rm c.m.}=1950$~MeV.
\label{fig:m_kk_mc_exp}}
\end{minipage}\hfill\hfill
\begin{minipage}[t]{0.46\textwidth}
	\centerline{\includegraphics[width=0.98\textwidth]{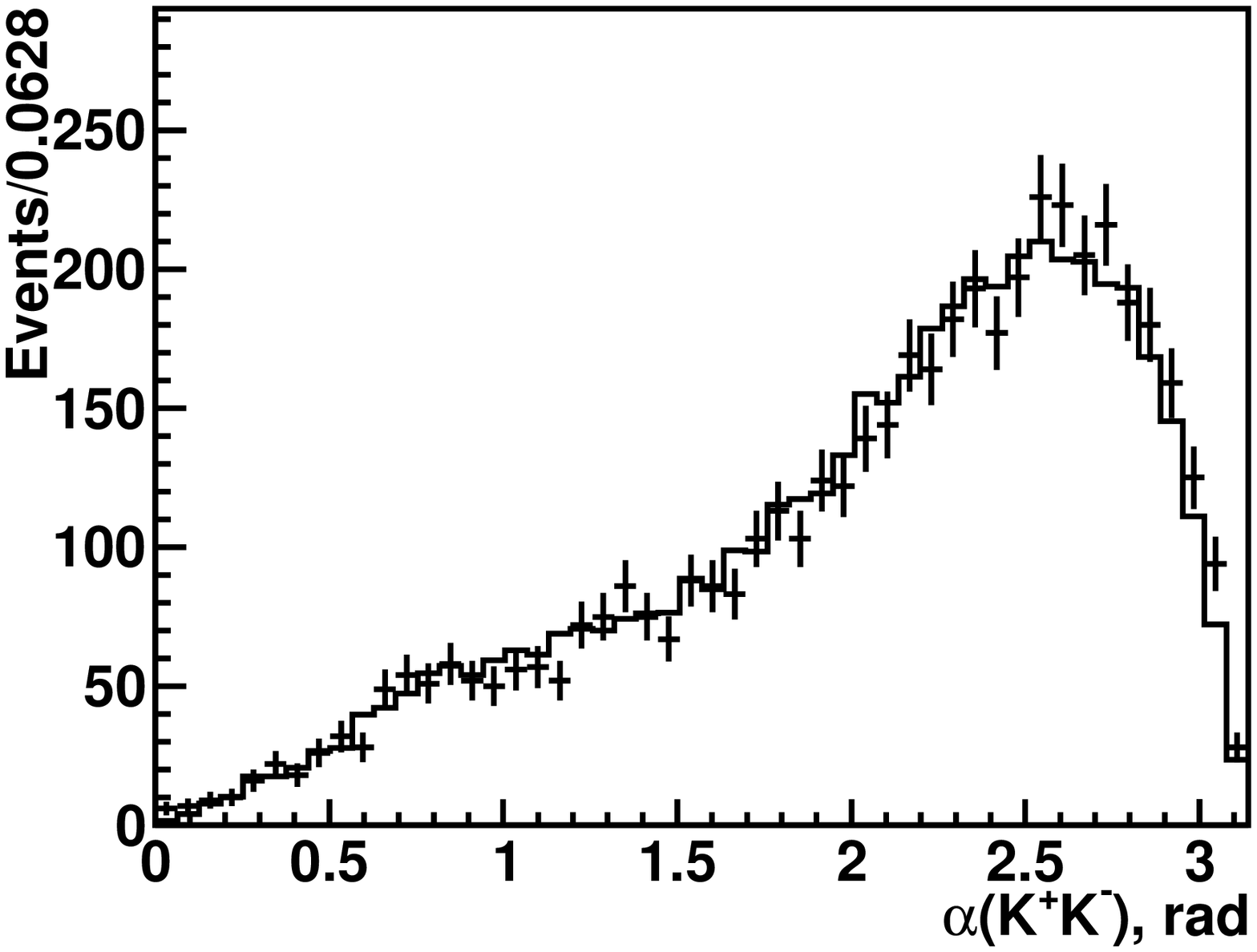}}
\caption
{The angle between $K^+$ and $K^-$ for data (points) and MC simulation 
(histogram) at $E_{\rm c.m.}=1950$~MeV.
\label{fig:ang_kk_mc_exp}}
\end{minipage}\hfill\hfill
\end{figure}
\begin{figure}[hbtp]
\begin{minipage}[t]{0.46\textwidth}
	\centerline{\includegraphics[width=0.98\textwidth]{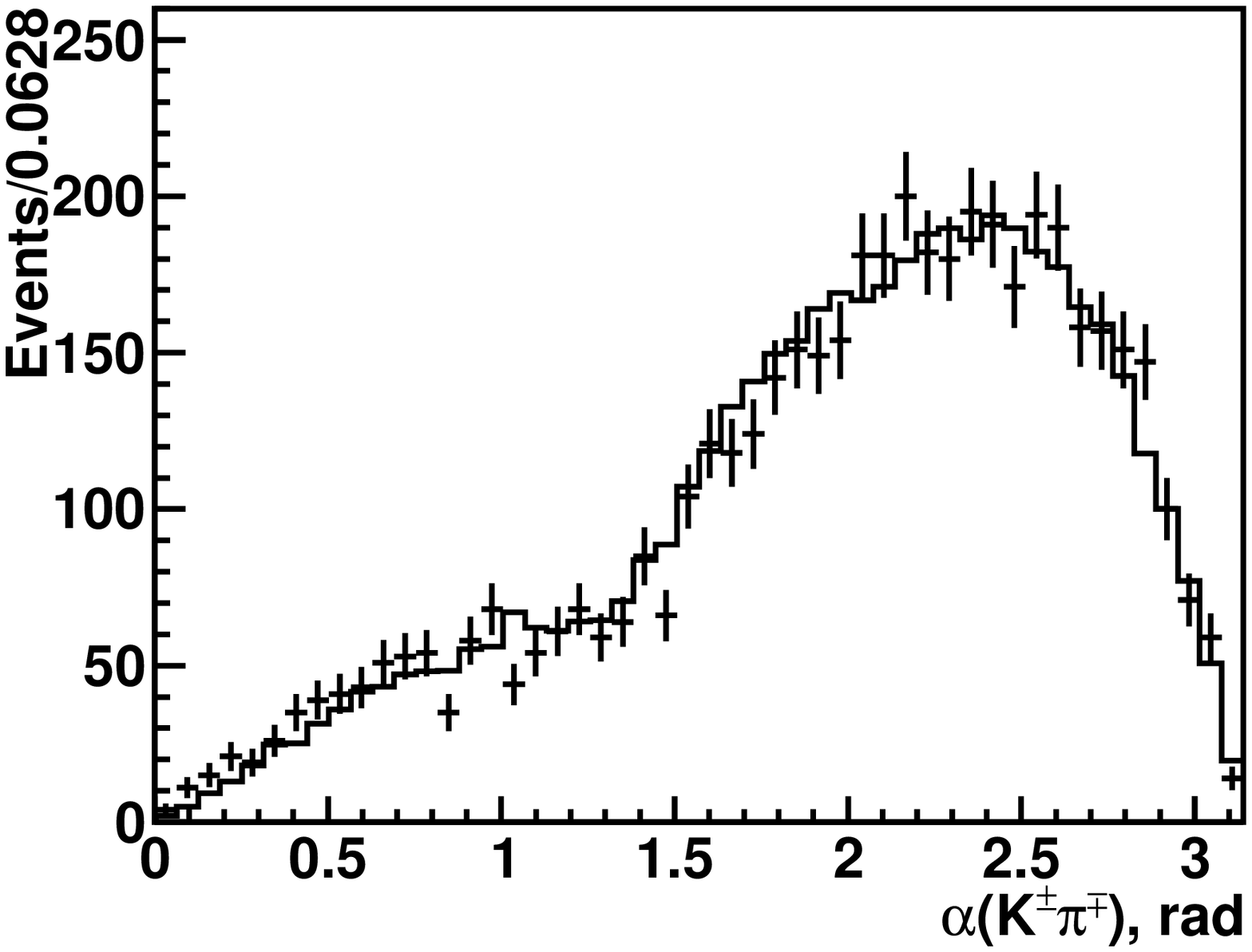}}
\caption
{The angle between $K^{\pm}$ and $\pi^{\mp}$ for data (points) and MC simulation 
(histogram) at $E_{\rm c.m.}=1950$~MeV.
\label{fig:ang_kpi_mc_exp}}
\end{minipage}\hfill\hfill
\begin{minipage}[t]{0.46\textwidth}
	\centerline{\includegraphics[width=0.98\textwidth]{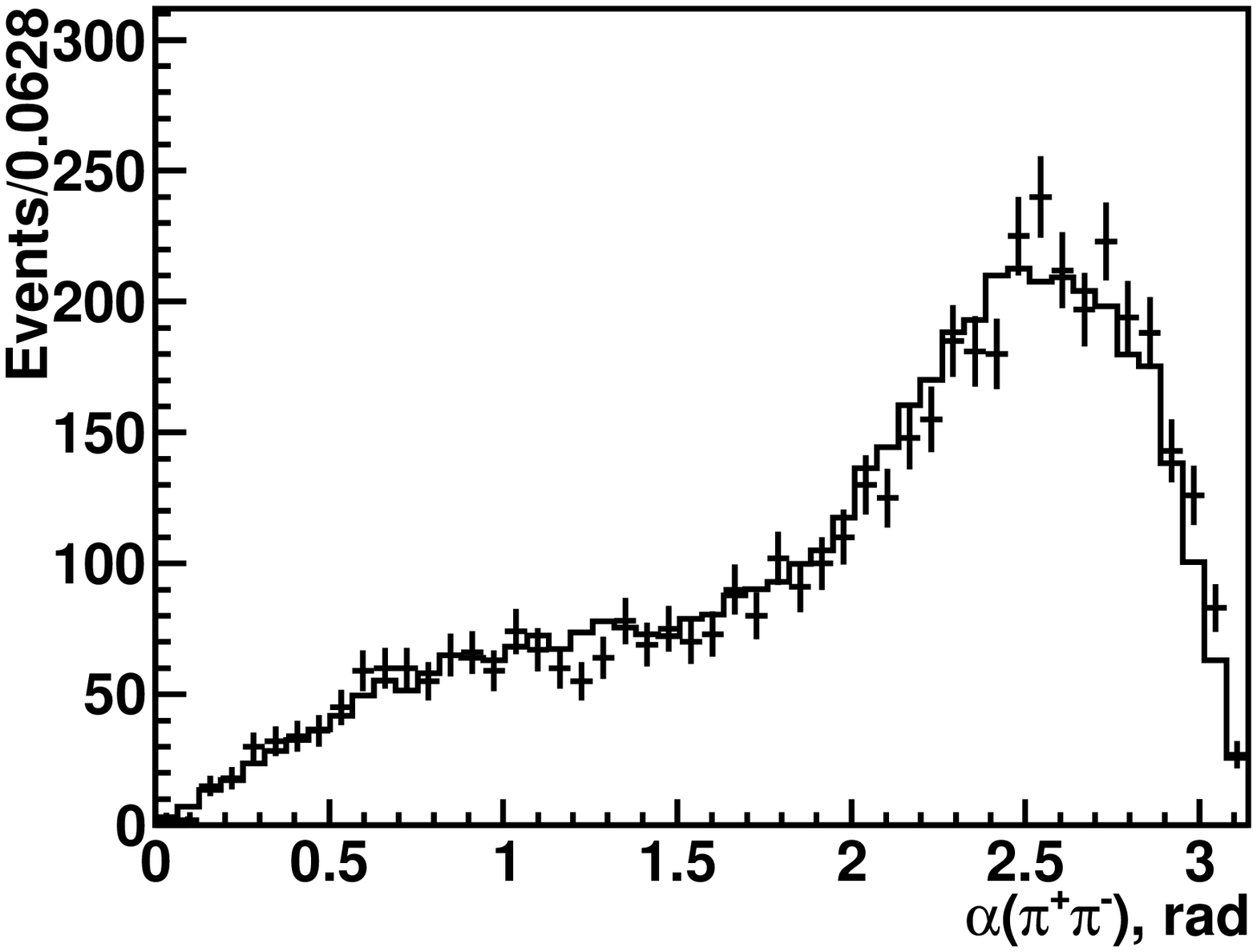}}
\caption
{The angle between $\pi^+$ and $\pi^-$ for data (points) and MC simulation 
(histogram) at $E_{\rm c.m.}=1950$~MeV.
\label{fig:ang_pipi_mc_exp}}
\end{minipage}\hfill\hfill
\end{figure}

Comparison of simulated and experimental invariant mass and angular 
distributions for the c.m. energy of 1950\,MeV is shown in 
Figs.~\ref{fig:m_kpi_mc_exp}--\ref{fig:ang_pipi_mc_exp}.
The points with error bars correspond to experimental data, the histograms 
correspond to simulation; the obtained mixture of the intermediate states 
reasonably describes the data.

\section{Efficiency\label{sec:efficiency}}
\begin{figure}[hbtp]
\begin{minipage}[t]{0.46\textwidth}
	\centerline{\includegraphics[width=0.98\textwidth]{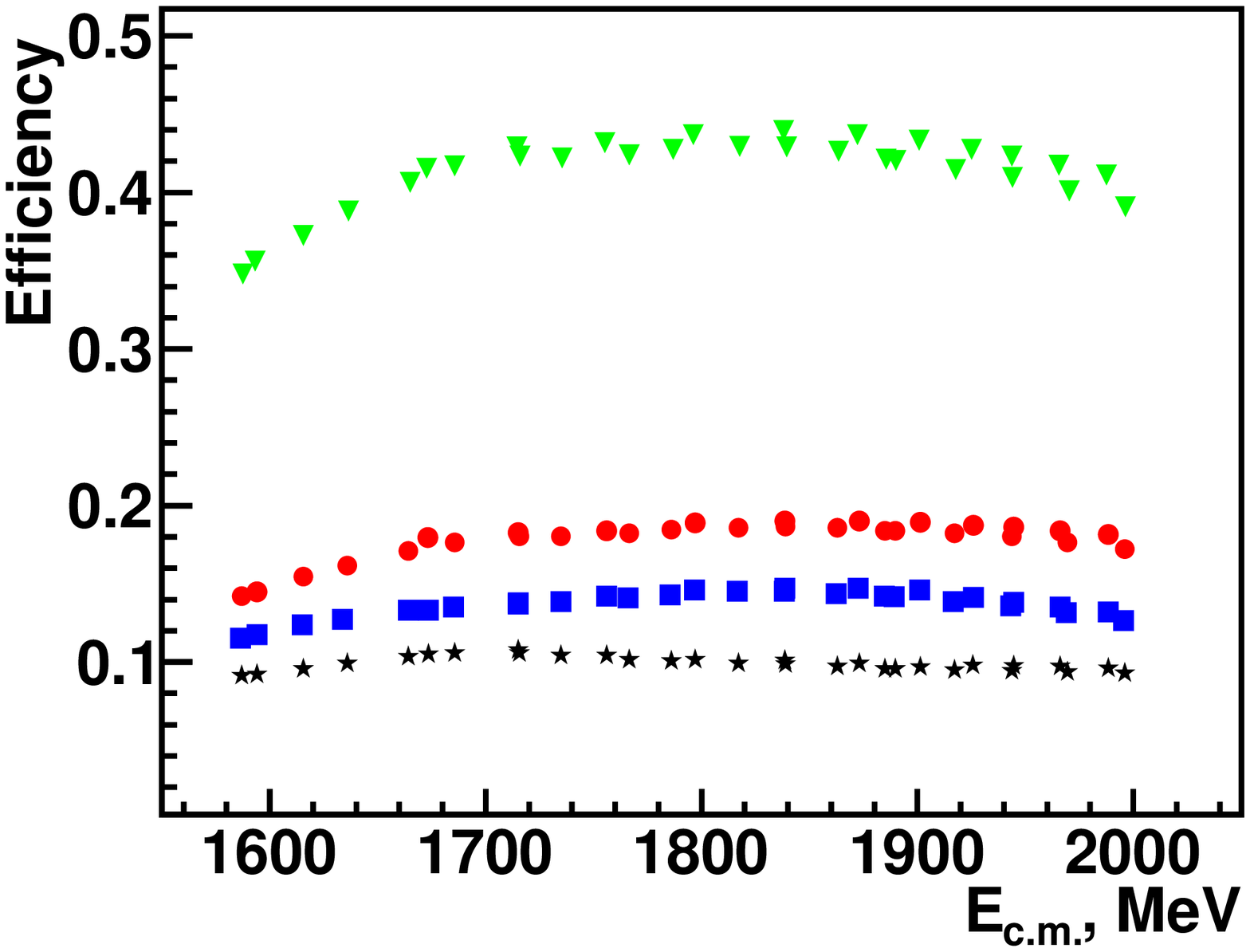}}
\caption
{Detection efficiency for four-track events (circles), events with a missing 
pion (stars), events with a missing kaon (squares) and total detection 
efficiency (triangles) vs c.m. energy.
\label{fig:eff}}
\end{minipage}\hfill\hfill
\begin{minipage}[t]{0.46\textwidth}
	\centerline{\includegraphics[width=0.98\textwidth]{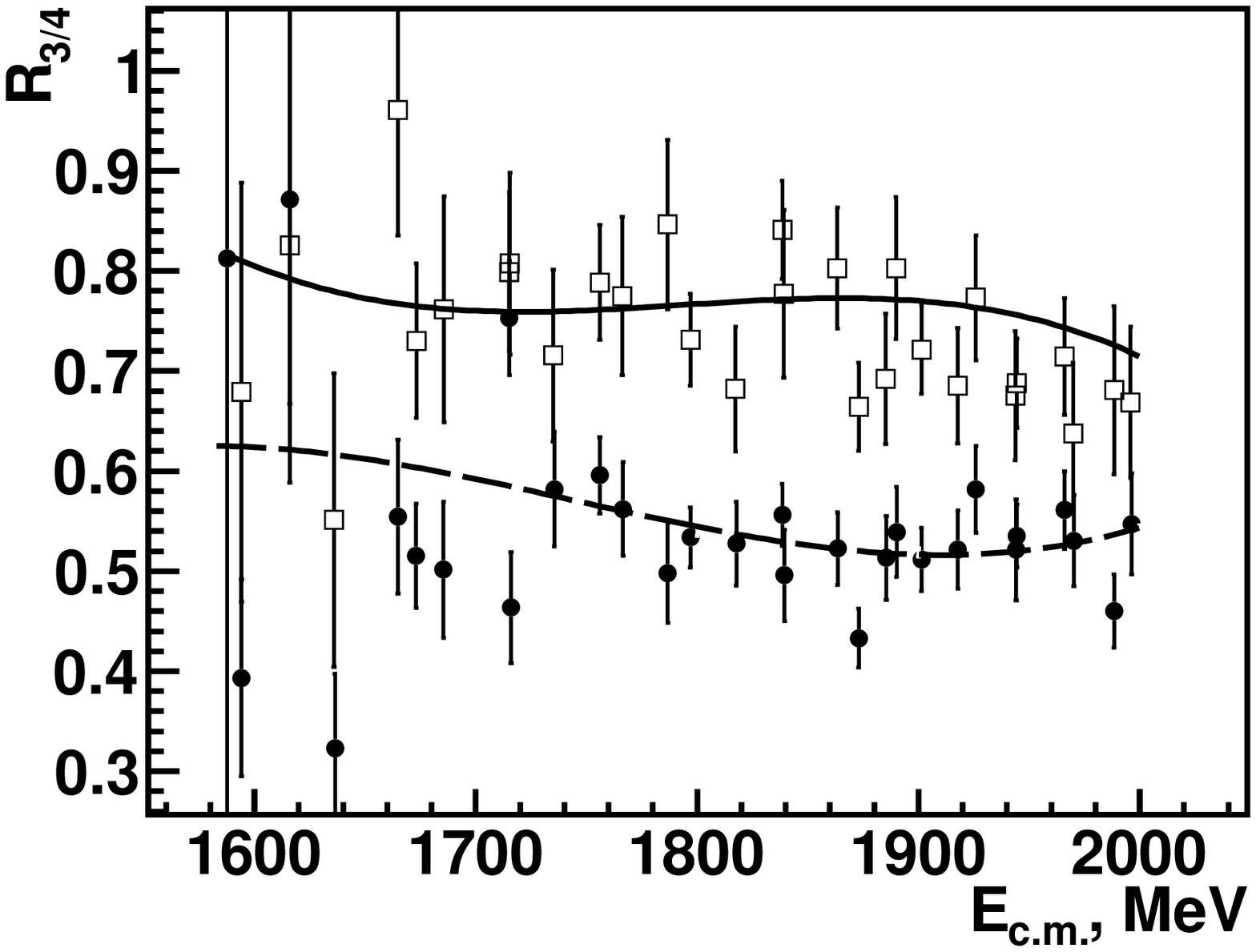}}
\caption
{The ratio of three- and four-track events, $R_{3/4}$, for a missing 
pion (circles) and a missing kaon (squares). The lines show corresponding 
simulated values.
\label{fig:n34}}
\end{minipage}\hfill\hfill
\end{figure}

We calculate the detection efficiency from simulation as a ratio of the
number of events after the selections described in Sec.~\ref{sec:selections} 
and the total number of simulated events. 
Figure~\ref{fig:eff} shows the detection efficiency calculated in our model 
for four-track events (circles), events 
with a missing pion (stars), events with a missing kaon (squares), and 
the total one (triangles) vs c.m. energy. 
The detection efficiency calculated in our model is 10\% smaller than that
in a phase space model.

Since events with a missing track are mostly due to the limited DC acceptance, 
the ratio of the number of three-track and four-track events 
$R_{3/4}$ is sensitive to the polar-angle distributions, and provides 
an additional test of the production mechanism used in the MC simulation.
Figure~\ref{fig:n34} shows $R_{3/4}$ for missing pions (circles) and 
missing kaons (squares) in comparison with the corresponding simulated values 
shown by the smooth lines. We observe good agreement of the data and simulation
and use this comparison to estimate systematic uncertainties on the 
detection efficiency discussed below. 

Three-track events with a missing pion or kaon expected in the DC acceptance 
are used to estimate the single-track DC reconstruction efficiency for data 
and MC simulation.
It is calculated as 
$\varepsilon_{1tr} = 2/(2 + R_{3/4})$, and we obtain 
$\varepsilon_{1tr~K}^{\rm exp} = 0.855 \pm 0.005$ 
($\varepsilon_{1tr~K}^{\rm MC} = 0.862 \pm 0.005$) and
$\varepsilon_{1tr~\pi}^{\rm exp} = 0.95 \pm 0.01$ 
($\varepsilon_{1tr~\pi}^{\rm MC} = 0.955 \pm 0.005$)
averaged over the c.m. energy range 1600--2000\,MeV for kaons and pions, respectively.

Using these values we calculate the corrections to the number of three- 
and four-track experimental events, $\xi_{4tr}$ and $\xi_{3tr}$: 
$$\xi_{4tr}=\frac{(\varepsilon_{1tr~K}^{\rm exp})^2  (\varepsilon_{1tr~\pi} ^{\rm exp})^2}{(\varepsilon_{1tr~K} ^{\rm MC})^2 (\varepsilon_{1tr~\pi} ^{\rm MC})^2},
$$
$$
\xi_{3tr} ^{\pi}=\frac{(\varepsilon_{1tr~K} ^{\rm exp})^2 \varepsilon_{1tr~\pi}^{\rm exp} (1 - \varepsilon_{1tr~\pi} ^{\rm exp})}{(\varepsilon_{1tr~K}^{\rm MC})^2 \varepsilon_{1tr~\pi}^{\rm MC} (1 - \varepsilon_{1tr~\pi}^{\rm MC}) },~~~
$$
$$
\xi_{3tr} ^{K}=\frac{(\varepsilon_{1tr~\pi} ^{\rm exp})^2 \varepsilon_{1tr~K}^{\rm exp} (1 - \varepsilon_{1tr~K} ^{\rm exp})}{(\varepsilon_{1tr~\pi}^{\rm MC})^2 \varepsilon_{1tr~K}^{\rm MC} (1 - \varepsilon_{1tr~K}^{\rm MC}) }.
$$
These corrections depend on  conditions used to determine  "good" tracks, 
and their values vary in the 0.95--1.05 range when 
we study systematic uncertainties (see below).
For selected "good" tracks our simulation describes well the DC detection 
efficiency and  corrections are about unity.

\begin{figure}[hbtp]
\begin{minipage}[t]{0.46\textwidth}
	\centerline{\includegraphics[width=0.98\textwidth]{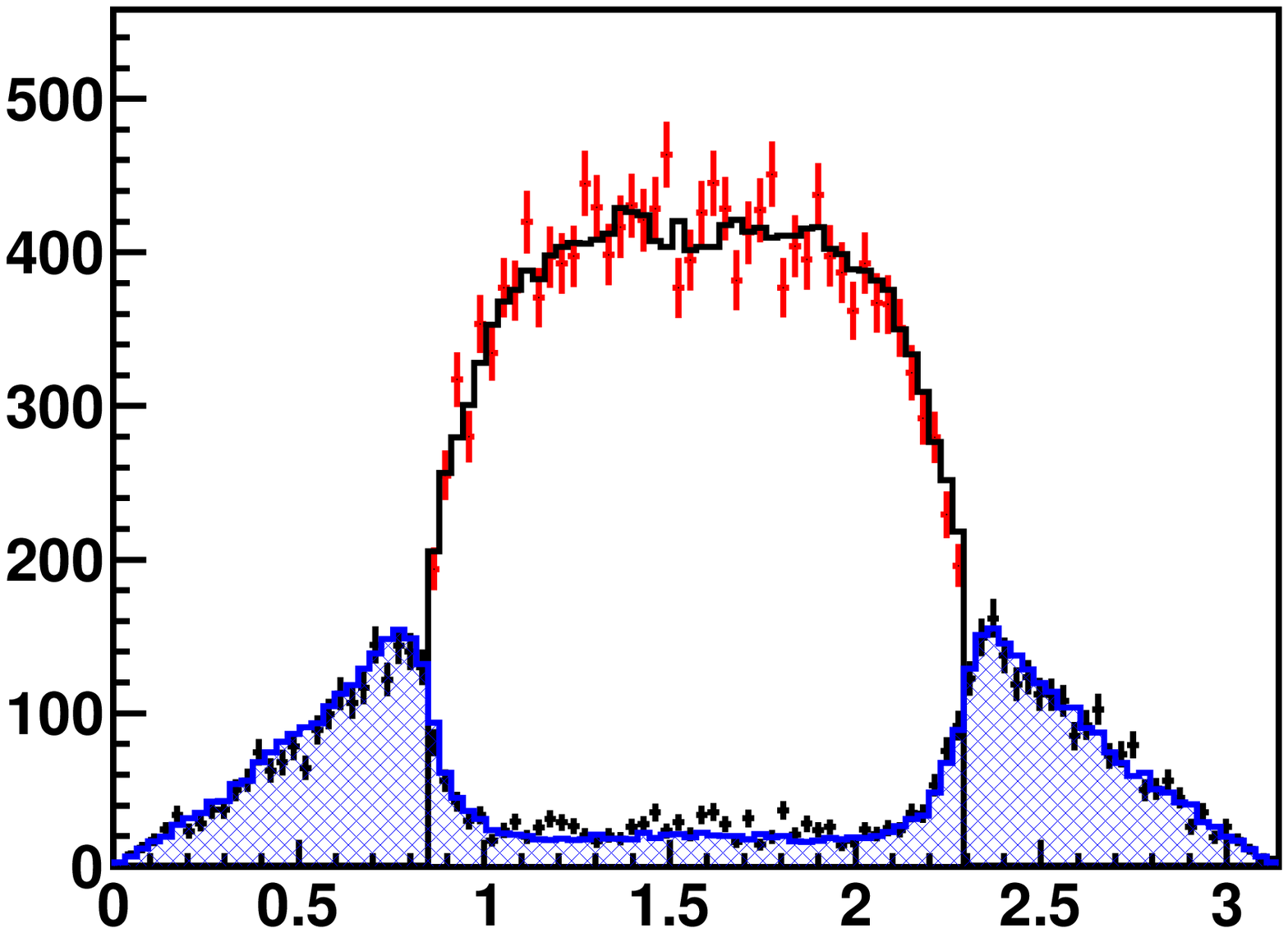}}
\caption
{The polar angle of kaon or pion. Points correspond to experimental data, 
	histogram -- simulation. The open region corresponds to the 
polar angle of detected particles,
	shaded -- polar angle of the particles that were not detected.
\label{fig:theta}}
\end{minipage}\hfill\hfill
\begin{minipage}[t]{0.46\textwidth}
	\centerline{\includegraphics[width=0.98\textwidth]{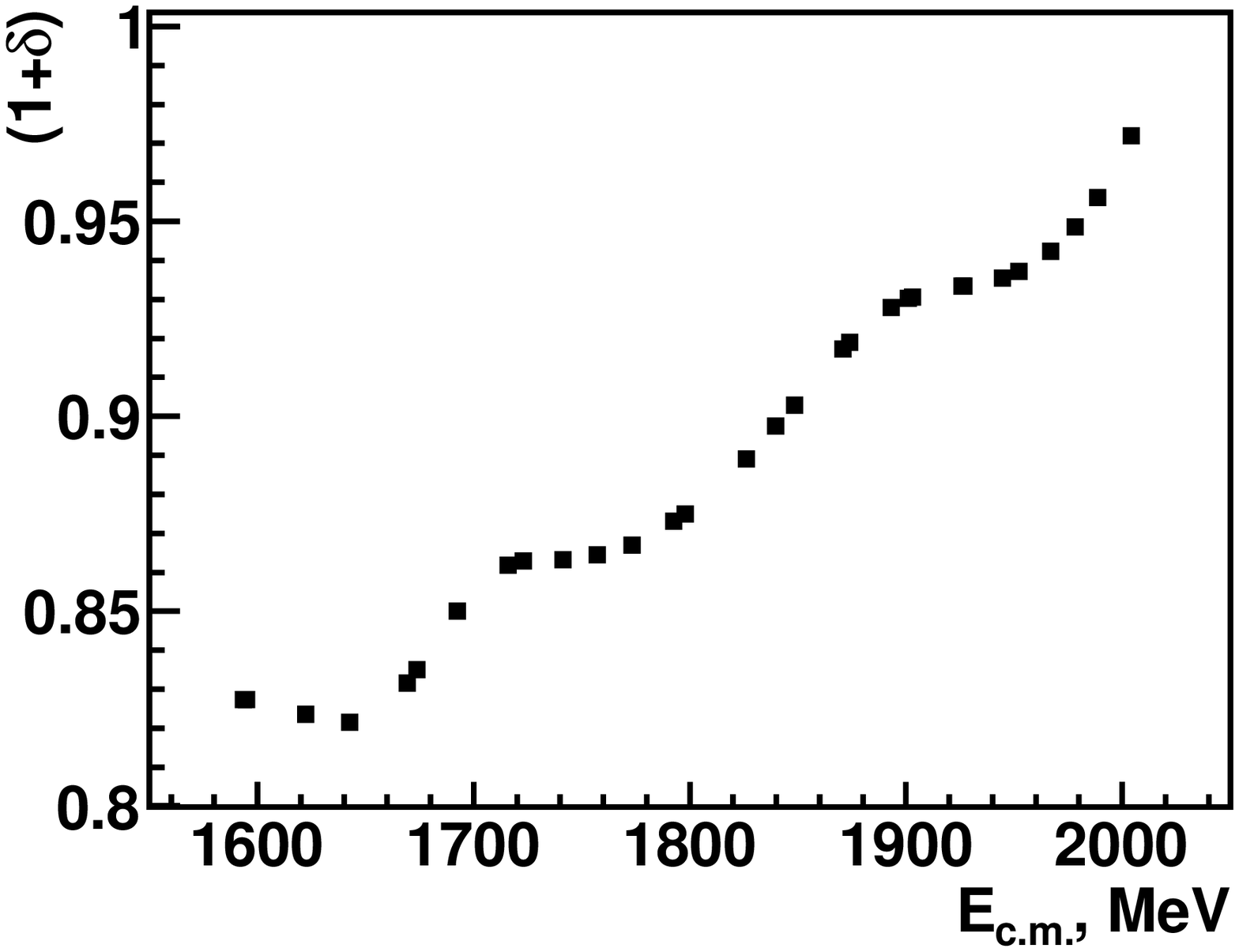}}
\caption
{A radiative correction versus c.m. energy.
\label{fig:rc}}
\end{minipage}\hfill\hfill
\end{figure}

Figure~\ref{fig:theta} presents the polar angle distribution for 
three-track events after background subtraction. 
The points correspond to experimental data, the open histogram is for 
simulation of tracks detected in DC, and the shaded histogram shows 
the distribution for the expected direction of missing particles. 
The simulation describes our data well.

\section{Cross Section Calculation}
At each energy point the Born cross section for the process 
$e^{+}e^{-} \to K^{+} K^{-} \pi^{+} \pi^{-}$ 
is calculated as a weighted average of the cross sections for three 
different data sets:
$$
\sigma_ {4tr} = \frac{ N_{4tr}}{\epsilon_{4tr} L_{int} (1+\delta) \xi_{4tr}},
$$
$$
\sigma_{3tr~K} = \frac{ N_{3tr~K}}{\epsilon_{3tr~K} L_{int} (1+\delta) \xi_{3tr~K}},
$$
and
$$
\sigma_ {3tr~\pi} = \frac{ N_{3tr~\pi}}{\epsilon_{3tr~\pi} L_{int} (1+\delta) \xi_{3tr~\pi}} 
$$
for four- and three-track events,
where $L_{\rm int}$ is the integrated luminosity, $(1+\delta)$ is the radiative 
correction,
$\epsilon_{4tr}$ and $\epsilon_{3tr~K,\pi}$ are 
the detection efficiencies from simulation, $\xi_{4tr}$ and $\xi_{3tr~K,\pi}$
are corrections for the data-MC difference in the DC track reconstruction 
efficiencies.

The radiative correction shown in Fig.~\ref{fig:rc} is calculated according 
to~\cite{kur_fad,rmc}, using our cross section data and an iteration procedure.

\begin{figure}[hbtp]
\begin{minipage}[t]{0.46\textwidth}
	\centerline{\includegraphics[width=0.98\textwidth]{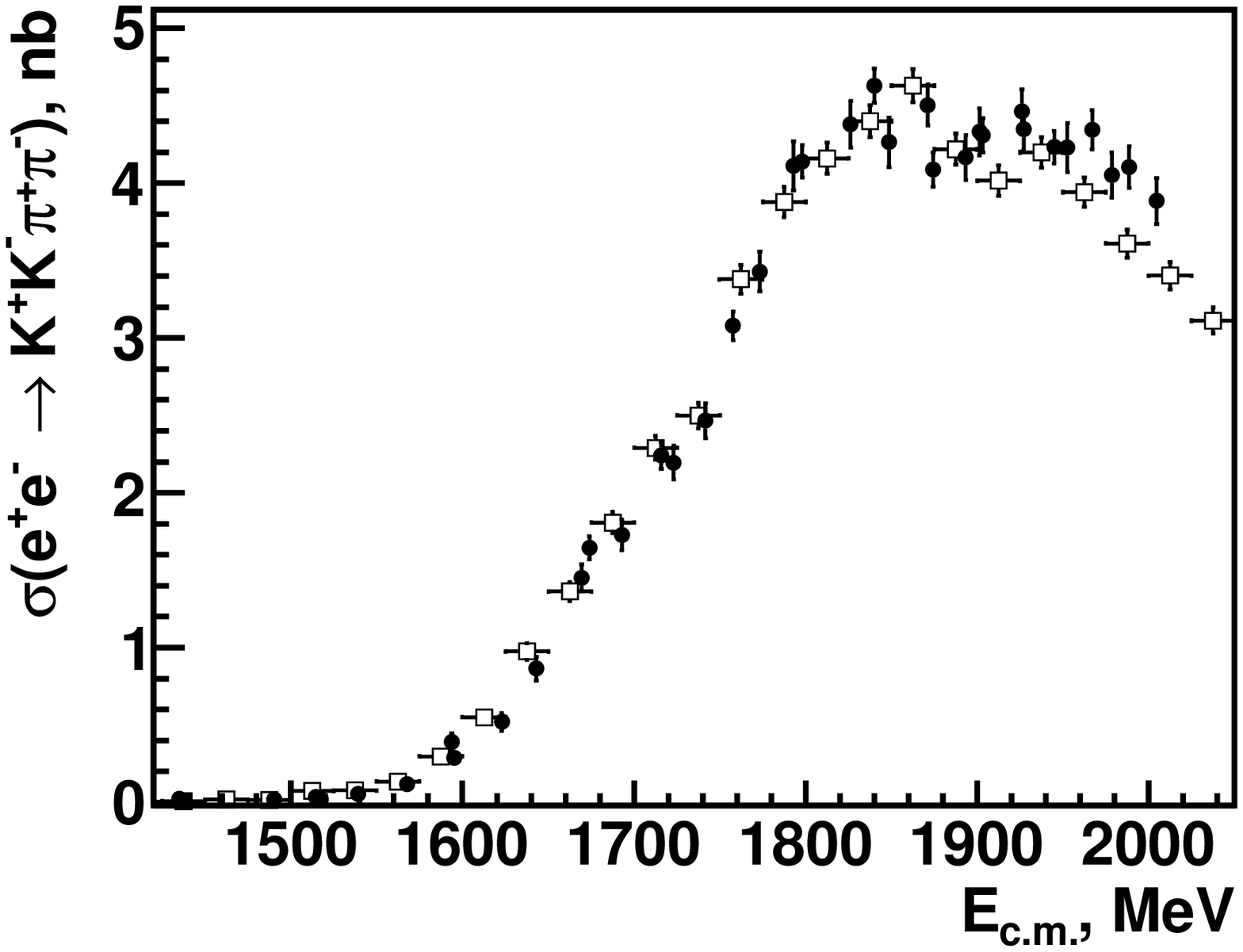}}
\caption
{The cross section of the process $e^{+}e^{-} \to K^{+} K^{-} \pi^{+} \pi^{-}$ 
obtained with the CMD-3 detector (dark circles) in comparison with 
the BaBar measurement (open circles).
\label{fig:cs_all}}
\end{minipage}\hfill\hfill
\begin{minipage}[t]{0.46\textwidth}
	\centerline{\includegraphics[width=0.98\textwidth]{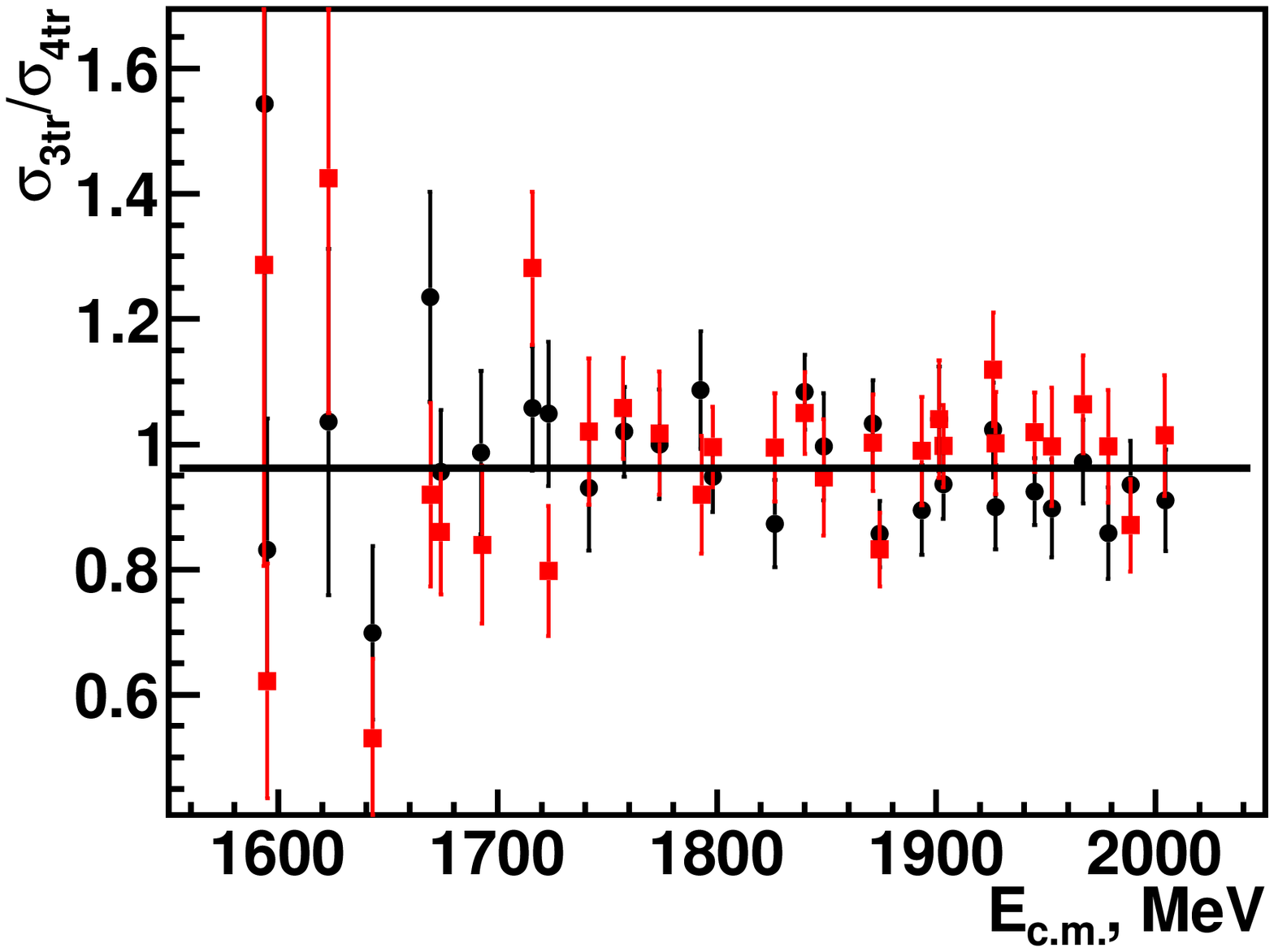}}
\caption
{The ratio of three-track and four-track cross sections. The line is an 
approximation.
\label{fig:cs_3_4}}
\end{minipage}\hfill\hfill
\end{figure}

The obtained cross section is presented in 
Fig.~\ref{fig:cs_all} as a function of c.m. energy.
Our result agrees with the previous measurement performed
by the BaBar Collaboration~\cite{babar2} and has comparable or better 
statistical precision.

The integrated luminosity, the number of four- and three-track events, 
detection efficiency, radiative correction and obtained cross section for 
each energy point are listed in Table 1. Only statistical errors are shown.

\section{Systematic errors\label{sec:systematic}}
The following main sources of systematic uncertainties are considered:
\begin{itemize}
\item The uncertainty on the determination of the integrated luminosity 
is 1\%~\cite{lum};
\item A systematic error due to the $K/\pi$ separation is estimated to be 0.5\%;
\item Using the ratio of the cross section values obtained for the 
2011 and 2012 runs we estimate a  
systematic uncertainty associated with overall detector operation as  1.2 \%;
\item The accuracy of background subtraction for three-track events is
	studied by the variation of the fit range and 
is estimated as 1.5\%;
\item A systematic error due to the selection criteria is studied by varying
the "good" track definition  and is estimated to be 2.6\%;
\item The model dependence uncertainty of 5\% is conservatively estimated by varying 
the polar angle selection, and by comparison of the cross sections for 
four- and three-track events shown in Fig.~\ref{fig:cs_3_4}.
The mean value of the ratio of such events is 0.960$\pm$0.024.
\end{itemize}

The above systematic uncertainties summed in quadrature give an overall
systematic error of  6.1\%.

An additional systematic uncertainty is related to the
accuracy of c.m. energy determination, which is about 6\,MeV and 2\,MeV 
for the 2011 and 2012 runs, respectively.
It leads to an energy-dependent uncertainty of about 10\% at 
$E_{\rm c.m.}$=1500-1600\,MeV linearly decreasing to 1\% at $E_{\rm c.m.}$=1800\,MeV.
In the energy range $E_{\rm c.m.}$=1800---2000\,MeV this uncertainty is less 
than 1\%. The c.m. energy uncertainties are listed in 
Table~\ref{tab:cs_kkpipi1}.

\section{Conclusion}
The total cross section of the process $e^+e^- \to K^+K^-\pi^+\pi^-$ has 
been measured using 23 pb$^{-1}$
of an integrated luminosity collected with the CMD-3 detector at the 
VEPP-2000  collider in the 1500 - 2000\,MeV c.m. energy range. 
Our results agree with the previous measurements and have comparable or better 
statistical precision.
This final state exhibits complex resonant substructures.
Our tentative study of dynamics shows the following major intermediate 
mechanisms: 
$e^{+}e^{-} \to (K_1(1270) K)_{\rm S-wave} \to (K^* \pi)_{\rm S-wave} K$,
$e^{+}e^{-} \to (K_1(1400) K)_{\rm S-wave} \to (K^* \pi)_{\rm S-wave} K$,
$e^{+}e^{-} \to (K_1(1270) K)_{\rm S-wave} \to (\rho K)_{\rm S-wave}K$,
$e^{+}e^{-} \to f_0(980)\phi$,
$e^{+}e^{-} \to f_0(500) \phi$
and $e^{+}e^{-} \to \rho (KK)_{\rm S-wave}$.
Simulation based on these models is in good agreement with the 
experimental data and allows one to measure the total cross section with 
a systematic uncertainty falling from 11.7\% at 1500-1600~MeV to 6.1\% above
1800\,MeV.
After VEPP-2000 upgrade it will be possible to perform a more detailed analysis 
of dynamics using higher statistics and various charge combinations
of the $K\bar{K}\pi\pi$ final state. We also plan to improve $K/\pi$ separation 
by using additionally ionization losses in the xenon barrel calorimeter. 

\section{Acknowledgment}
This work is supported in part by the RFBR grants 
13-02-00215, 13-02-01134, 14-02-31275, 14-02-00047, 14-02-91332, 14-02-31478,
15-02-05674 and the DFG grant HA 1457/9-1.
Investigation of the ionization losses in the LXe calorimeter 
(section~\ref{sec:selections})
has been supported by Russian Science Foundation (Project 14-50-00080).

\begin{table}[hbtp]
	\caption{\label{tab:cs_kkpipi1}
		Center-of-mass energy, integrated luminosity, number of 
four-track events, number of three-track events, detection efficiency,
	radiative correction and Born cross section of the process 
$e^{+}e^{-} \to K^{+} K^{-} \pi^{+} \pi^{-}$. Errors are statistical only. }
\begin{center}
\begin{tabular}{c c c c c c c}
\hline
E$_{\rm c.m.}$,\,MeV & Lum., nb$^{-1}$  &  N$_{4}$ & N$_{3}$ & $\epsilon$ & 
$(1+\delta)$ & $ \sigma$, nb \\ 
\hline

 2004.6 $\pm$ 6 & 478.1 &  313 & 380.5 & 0.390 & 0.972 & 3.88 $\pm$ 0.15 \\ 
 1988.6 $\pm$ 2 & 600.6 &  445 & 507.7 & 0.410 & 0.956 & 4.10 $\pm$ 0.13 \\ 
 1978.4 $\pm$ 6 & 506.6 &  353 & 412.0 & 0.398 & 0.949 & 4.05 $\pm$ 0.15 \\ 
 1966.9 $\pm$ 2 & 692.2 &  510 & 650.6 & 0.416 & 0.942 & 4.35 $\pm$ 0.13 \\ 
 1952.6 $\pm$ 6 & 451.0 &  326 & 390.0 & 0.407 & 0.937 & 4.23 $\pm$ 0.16 \\ 
 1944.8 $\pm$ 2 & 993.8 &  735 & 898.9 & 0.422 & 0.935 & 4.23 $\pm$ 0.10 \\ 
 1927.0 $\pm$ 6 & 590.8 &  441 & 532.2 & 0.413 & 0.933 & 4.35 $\pm$ 0.14 \\ 
 1926.0 $\pm$ 6 & 566.9 &  420 & 569.1 & 0.426 & 0.933 & 4.47 $\pm$ 0.14 \\ 
 1903.2 $\pm$ 2 & 900.4 &  682 & 841.0 & 0.431 & 0.930 & 4.31 $\pm$ 0.11 \\ 
 1901.3 $\pm$ 6 & 498.6 &  351 & 471.0 & 0.418 & 0.930 & 4.33 $\pm$ 0.15 \\ 
 1893.4 $\pm$ 6 & 527.1 &  381 & 459.1 & 0.420 & 0.928 & 4.17 $\pm$ 0.14 \\ 
 1874.2 $\pm$ 2 & 855.6 &  659 & 723.0 & 0.436 & 0.919 & 4.09 $\pm$ 0.11 \\ 
 1871.1 $\pm$ 6 & 671.0 &  497 & 658.3 & 0.424 & 0.917 & 4.50 $\pm$ 0.13 \\ 
 1848.6 $\pm$ 6 & 435.4 &  311 & 395.8 & 0.429 & 0.903 & 4.27 $\pm$ 0.16 \\ 
 1840.0 $\pm$ 2 & 966.0 &  721 & 1007.6 & 0.438 & 0.897 & 4.63 $\pm$ 0.11 \\ 
 1826.4 $\pm$ 6 & 513.8 &  383 & 463.0 & 0.429 & 0.889 & 4.38 $\pm$ 0.15 \\ 
 1798.0 $\pm$ 2 & 998.4 &  684 & 865.4 & 0.436 & 0.875 & 4.14 $\pm$ 0.11 \\ 
 1792.9 $\pm$ 6 & 449.1 &  289 & 388.6 & 0.428 & 0.873 & 4.11 $\pm$ 0.16 \\ 
 1773.7 $\pm$ 6 & 560.6 &  297 & 396.9 & 0.425 & 0.867 & 3.43 $\pm$ 0.13 \\ 
 1757.7 $\pm$ 2 & 971.9 &  459 & 635.5 & 0.431 & 0.864 & 3.08 $\pm$ 0.09 \\ 
 1741.6 $\pm$ 6 & 542.3 &  208 & 269.9 & 0.422 & 0.863 & 2.47 $\pm$ 0.11 \\ 
 1723.1 $\pm$ 6 & 530.7 &  185 & 235.0 & 0.423 & 0.863 & 2.20 $\pm$ 0.11 \\ 
 1715.8 $\pm$ 2 & 812.1 &  257 & 398.8 & 0.428 & 0.862 & 2.24 $\pm$ 0.09 \\ 
 1692.8 $\pm$ 6 & 494.2 &  132 & 166.6 & 0.421 & 0.850 & 1.73 $\pm$ 0.10 \\ 
 1674.1 $\pm$ 2 & 894.7 &  220 & 273.5 & 0.415 & 0.835 & 1.64 $\pm$ 0.07 \\ 
 1669.4 $\pm$ 6 & 572.2 &  111 & 168.1 & 0.409 & 0.832 & 1.45 $\pm$ 0.09 \\ 
 1643.0 $\pm$ 6 & 462.7 &   72 & 62.9 & 0.393 & 0.821 & 0.87 $\pm$ 0.08 \\ 
 1622.9 $\pm$ 6 & 517.8 &   31 & 52.6 & 0.378 & 0.824 & 0.52 $\pm$ 0.06 \\ 
 1595.0 $\pm$ 2 & 832.7 &   39 & 35.8 & 0.356 & 0.827 & 0.29 $\pm$ 0.03 \\ 
 1593.8 $\pm$ 6 & 449.8 &   16 & 35.2 & 0.354 & 0.827 & 0.39 $\pm$ 0.06 \\ 
 1571.9 $\pm$ 6 & 522.0 &    7 & -    & 0.136 & 0.825 & 0.119 $\pm$ 0.045 \\
 1543.2 $\pm$ 6 & 512.0 &    3 & -    & 0.128 & 0.823 & 0.056 $\pm$ 0.032 \\
 1522.4 $\pm$ 6 & 539.5 &    1 & -    & 0.121 & 0.821 & 0.019 $\pm$ 0.019 \\
 1514.6 $\pm$ 2 & 847.4 &    3 & -    & 0.121 & 0.821 & 0.036 $\pm$ 0.020 \\
 1494.1 $\pm$ 6 & 556.8 &    1 & -    & 0.120 & 0.819 & 0.018 $\pm$ 0.018 \\
 1434.9 $\pm$ 6 & 927.3 &    2 & -    & 0.120 & 0.815 & 0.022 $\pm$ 0.016 \\

 \hline 
\end{tabular}
\end{center}
\end{table}

\end{document}